\documentclass{sig-alternate}

\usepackage[T1]{fontenc}
\usepackage{times}
\usepackage{helvet}
\usepackage{courier}
\usepackage{verbatim}
\usepackage{url}
\usepackage{bm}
\usepackage{color}
\usepackage{amsmath, amssymb,sansmath}
\usepackage{balance}
\usepackage{lipsum}
\usepackage{CJKutf8}
\usepackage[utf8]{inputenc} 
\usepackage[english]{babel}
\usepackage[autostyle]{csquotes} 

\makeatletter
\newif\if@restonecol
\makeatother

\usepackage[lined,algonl,boxed]{algorithm2e}
\usepackage{epsfig}
\usepackage{subfigure}
\usepackage{multirow}
\usepackage{amsmath}
\usepackage{color}

\usepackage{epstopdf}
\usepackage{tikz}
\usetikzlibrary{automata}
\graphicspath{{./}}
\newcommand{\hide}[1]{} 
\newcommand{\vpara}[1]{\vspace{0.1in}\noindent\textbf{#1 }}

\newcommand{\figref}[1]{Figure~\ref{#1}} 

\newdef{definition}{Definition}

\usepackage[
  pass,
]{geometry}

\hide{
\newfont{\mycrnotice}{ptmr8t at 7pt}
\newfont{\myconfname}{ptmri8t at 7pt}
\let\crnotice\mycrnotice%
\let\confname\myconfname%
}

\permission{Copyright is held by the International World Wide Web Conference Committee
(IW3C2). IW3C2 reserves the right to provide a hyperlink to the
author’s site if the Material is used in electronic media.}
\conferenceinfo{WWW}{'16, April 11-15, 2016, Montr\'eal, Qu\'ebec, Canada.}
\copyrightetc{ACM \the\acmcopyr}
\crdata{978-1-4503-4143-1/16/04.\\
	http://dx.doi.org/10.1145/2872427.2882979}

\begin{document}

\title{
The Lifecycle and Cascade of WeChat Social Messaging Groups
}

%

\numberofauthors{1} 
%

\author{
	%
	%
	\alignauthor
	Jiezhong Qiu$^{\dag}$, Yixuan Li$^{\sharp}$, Jie Tang$^{\dag}$,  Zheng Lu$^{\ddag}$, Hao Ye$^{\ddag}$, Bo Chen$^{\ddag}$, Qiang Yang$^{\star}$, \\and John E. Hopcroft$^{\sharp}$\\
	\affaddr{$^{\dag}$Department of Computer Science and Technology, Tsinghua University}\\
	\affaddr{$^{\sharp}$Department of Computer Science, Cornell University}\\
	\affaddr{$^{\ddag}$Tencent Corporation, Shenzhen, China}\\
	\affaddr{$^{\star}$Department of Computer Science, Hong Kong University of Science and Technology}\\
	\email{\sf qjz12@mails.tsinghua.edu.cn, \{yli,jeh\}@cs.cornell.edu, jietang@tsinghua.edu.cn} 
}
\hide{
\author{
%
%
\alignauthor
Jiezhong Qiu \\
\affaddr{Tsinghua University}\\
 \email{qjz12@mails.tsinghua.edu.cn}
\alignauthor
Yixuan Li \\
\affaddr{Cornell Unversity}\\
\email{yli@cs.cornell.edu}
\alignauthor Jie Tang\\
       \affaddr{Tsinghua University}\\
       \email{ jietang@tsinghua.edu.cn }\\
\and  
\alignauthor
Qiang Yang \\
	\affaddr{Hong Kong University of Science and Technology}\\
	\email{qyang@cse.ust.hk}
\alignauthor
John E. Hopcroft \\
      \affaddr{Cornell University}\\
      \email{jeh@cs.cornell.edu}
}

\additionalauthors{
	Zheng Lu, Hao Ye, Bo Chen and Yufei Zheng~(Tencent Inc., email:
	{\texttt{\{zhakeberglu, dariaye, jennychen, yufeizheng\}@tencent.com}})
}
}

\maketitle
\sloppy
\begin{abstract}
Social instant messaging services are emerging as a transformative form with which people connect, communicate with friends in their daily life --- they catalyze the formation of social groups, and they bring people stronger sense of community and connection.
However, research community still knows little about the formation and evolution of groups in the context of social messaging --- their lifecycles, the change in their underlying structures over time, and the diffusion processes by which they develop new members. 


In this paper, we analyze the daily usage logs from WeChat group messaging platform --- the largest standalone messaging communication service in China --- with the goal of understanding the processes by which {\em social messaging groups} come together, grow new members, and evolve over time. Specifically, we discover a strong dichotomy among groups in terms of their lifecycle, and develop a separability model by taking into account a broad range of group-level features, showing that long-term and short-term groups are inherently distinct. We also found that the lifecycle of messaging groups is largely dependent on their social roles and functions in users’ daily social experiences and specific purposes. Given the strong separability between the long-term and short-term groups, we further address the problem concerning the early prediction of successful communities. 

In addition to modeling the growth and evolution from group-level perspective, we investigate the individual-level attributes of group members and study the diffusion process by which groups gain new members. By considering members' historical engagement behavior as well as the local social network structure that they embedded in, we develop a membership cascade model and demonstrate the effectiveness by achieving AUC of 95.31\% in predicting inviter, and an AUC of 98.66\% in predicting invitee.    
\end{abstract}




\vspace{3em}
\noindent \textbf{\large Keywords}

\noindent social messaging; online community; group formation; information diffusion

\section{introduction}

The advent and proliferation of social instant messaging services have been shaping and transforming the way people connect, communicate with individuals or groups of friends, bringing users diverse and ubiquitous social experiences that traditional text-based short message service (SMS) could not. 
For example, WhatsApp is the most globally popular messaging service with more than 900 million monthly active users (MAUs), WeChat, the largest messaging service in China, has more than 600 million MAUs.
These tools have enriched the way people interact by including images, video, location information, audio and text messages. 
More importantly, they have also catalyzed the formation of social groups, bringing people a stronger sense of community and connection compared with traditional text messaging~\cite{church2013s}.

While past work has extensively studied the dynamics of group formation and evolution, much of the work is limited to the setting of online communities embedded within the social networking sites --- which is inherently different from groups seen in the context of social messaging. Previous study~\cite{church2013s} has shown that, for most social messaging tools adopters, the creation and use of instant group messaging occurs more frequently and habitually than other form of group-level social engagement in their daily life. In terms of lifecycle, social messaging groups have a relatively shorter life span --- ranging from several hours to months --- as opposed to those online groups seen in social networking sites such as Reddit \cite{buntain2014identifying} and Facebook \cite{park2009being} that can sustain up to years. Furthermore, all the chat groups are by default only visible to the group members and grow in an invitation-only fashion, i.e., new members invited to the group are guaranteed to be on the fringe of group networks (one-hop neighbors of current group members) --- thus the membership cascade process is more locally dependent, with unidirectional contagion dominated mostly by the existing group members. This is very dissimilar from the diffusion and growing models in previous literature on online communities (e.g., \cite{backstrom2006group,palla2007quantifying}), in which users can make their own decisions to join, even if they are not friends with any of the current group members. 

Researchers have recently begun interpreting the group messaging behavior and processes from a social science perspective, yet concrete empirical measurement and statements cannot be drawn from existing literature.    Much of the challenge has been the lack of appropriate datasets --- one needs a large collection of messaging groups with sufficient time-resolution so that one can keep track of their emergence, growth and demise over time. Another challenge comes from devising an effective model to depict and quantify the diversified, complex processes by which the groups develop over time. As a result, the research community still knows very little about the formation and evolution of chat groups in the context of social messaging --- their lifecycles, the change in their underlying structures over time, and the cascade processes by which they develop new members. 




To address these issues, in this paper, we analyze the daily usage logs from the WeChat \footnote{\url{www.wechat.com/en/}} group messaging platform --- the largest standalone messaging communication service developed by Tencent in China \cite{wiki} --- with the goal of understanding the  processes by which social messaging groups come together, grow new members, and evolve over time. To our knowledge, this is so far the first large-scale analysis on messaging group dynamics. 
 WeChat allows users to send and receive multimedia messages in real-time via Internet. One important feature in WeChat is that any user can create a new group and invite friends to join this group. Please note that such group is invited only, which means that the other users (friends) cannot apply to join if no invitation comes from the group. 
 Groups play a very important role in WeChat. Our statistics show that roughly 25\% of the messages in WeChat were generated in group conversations. On the other hand, the groups are very dynamic. Every day, about 2,300,000 new groups were created  and about 40\% of the newly created groups become silent within only one week.
 We will describe detailed information about the WeChat dataset along with its mechanics in Section \ref{sec:data}.


\vspace{1em}
{\bf The Present Work: Lifecycle Dichotomy in Social Messaging Groups.}
In this paper, we contribute to research on group evolution in social messaging platforms by observing and making a conceptual difference between two types of groups in terms of their lifecycle: long-term and short-term groups. Our empirical analysis shows that almost 40\% of them stop interaction within one week. On the other hand,  we also observe 30\% of  the groups can survive a much longer period of time ($\ge 30$ days). The strong lifecycle dichotomy of chat groups leads us to a natural lifecycle modeling and prediction questions --- how separable are the long-term and short-term groups by taking into account the structural and social behavioral features? 
To address this issue, we develop a {\em separability model} by studying snapshots of millions of groups,  and show the strong distinction between long-term and short-term groups --- measured with a broad range of features including the underlying group network structure, the membership cascade tree properties (e.g. tree size and depth), and the demographics entropy of group members such as gender, age and region.   

We also discuss the phenomena of lifecycle dichotomy from the perspective of the roles and functions that social messaging platforms have in users' daily social experiences. This leads to the question of how does the lifecycle and growth pattern of social messaging groups correlate with the social functions it is serving? It turns out that messaging groups have been commonly adopted as a convenient way of connecting with smaller communities all at once, e.g., a family group, a colleague's group, a classmate's group, as well as groups for social events \cite{church2013s}. And the lifecycle of messaging groups is largely dependent on it social purpose for being setup --- for instance, one may expect that event-driven groups will have a higher chance of dying out than friend groups for frequent catching-up. 

Furthermore, given the strong separability between the long-term and short-term groups, a fundamental problem concerning the design of successful communities is: Can we predict whether a social group will grow and persist in the long run by analyzing the structural and behavioral patterns exhibited by the group at its early stage? We phrase it as a problem of {\em early prediction} in group longevity. Through the lens of various features exhibited by a group, we demonstrate that strong prediction results can be obtained even with a group history of one day. 
 
\vspace{1em}
 
{\bf The Present Work: Group Membership Cascade and Prediction.}
In addition to modeling the growth and evolution from a group-level perspective, we take one step further and investigate the individual-level attributes of group members and study the cascade process by which groups gain new members.  Specifically, given the historical behavior of group users as well as the local social structure, can we predict which users in the group are more likely to be active and invite new users to the group chat and to whom will he/she send invitations to? Making sense of such questions requires fine-grained  inspection into users' historical engagement behavior as well as the local social network structure that users embedded in. To this end, we develop a membership cascade process model in which we consider features of both {\em inviter} --- a group member who sends invitation to friend(s), and {\em invitee} --- the individual in the inviter's ego networks who gets invited to the group chat. Our inviter prediction model using all features generally achieves AUC as high as 95.31\%, and invitee prediction model reaches AUC of 98.66\%.

Furthermore, we also attempt to analyze: how does the added new members in return lead to the change of underlying social network structure, as the group evolves over time? To address this issue, we take snapshots and compare the same set of sample group at the timestamp of setup and after a month, respectively. Interestingly, we observe that although both long-term and short-term groups have increment on features such as close triads, long-term groups have shown to increase the close triads more significantly.


\vpara{Organization.} The remainder of the paper is organized as follows. Section \ref{sec:related} describes related work on analyzing group formation and evolution. In Section \ref{sec:data}, we introduce the WeChat social messaging group dataset. The discussion on group lifecycle dichotomy as well as early prediction model is provided in Section \ref{sec:dichotomy}. The membership cascade process is investigated in Section \ref{sec:cascade}. Finally we conclude our work in Section \ref{sec:conclusion}.


\section{RELATED WORK}
\label{sec:related}

The study of groups and communities is central to many research problems on mining and analytics of sociological data. There have been two major lines of research in this domain: one focuses on the static snapshots of social graphs and seeks to infer and identify tightly-connected group of members ---  also known as community detection in literature~\cite{girvan2002community, hopcroft2003natural, li2015uncovering, newman2004detecting,SunTang:13TKDE}; another line of research focusing on the group dynamics --- the growth and evolution of social groups --- is more related to our work here. Below we highlight a few and exlpain how ours contribute to the existing research.

\vpara{Group Dynamics.}To understand the process of how social groups form and evolve over time, previous work has extensively investigated the growth and longevity of various forms of online communities such as Facebook apps \cite{kloumann2015lifecycles}, game communities \cite{ducheneaut2007life}, knowledge sharing communities \cite{yang2010activity} and social networks communities \cite{backstrom2006group,butler2001membership,kairam2012life,palla2007quantifying,SunTang:13TKDE}. A more generalized work of Riberiro \cite{ribeiro2014modeling} investigates the group growth dynamics by encompassing a broad range of 22 membership-based websites.  

Our work focuses on studying the group dynamics of a rather understudied realm --- the social messaging services. Although researchers have recently begun interpreting the group messaging behavior from a social science perspective (e.g., \cite{church2013s}), research community still knows little about the formation and evolution of social messaging groups with concrete and empirical measurement. Focusing on the Yahoo! instant messaging traffic data, Aral et al. \cite{aral2009distinguishing} considers the individual-level dynamics from the perspective of peer influence and homophily, yet it is unclear how the groups as a whole evolve over time --- their lifecycles, their structural dynamics etc. Our work contributes to the current research by discovering a strong dichotomy among social messaging groups in terms of their lifecycle.  By taking into account a broad range of group-level structural and behavioral features, we develop a separability model that distinguishes between long-term groups and short-terms groups, as well as an early prediction model that forecasts the longevity of groups.

\vpara{Cascades.}The second half our of work on group membership cascade prediction builds on previous literature that studies diffusion processes \cite{gomez2010inferring}. In recent years, scientiests have been able to observe and quantify large-scale diffusions from the richness of online data, including blog space \cite{adar2004implicit,leskovec2007patterns}, marketing \cite{leskovec2007dynamics}, social sites such as LinkedIn \cite{anderson2015global}, Flickr \cite{cha2008characterizing}, Twitter \cite{goel2013structural,kwak2010twitter,romero2011differences}, Facebook \cite{bakshy2012role,cheng2014can,kloumann2015lifecycles, sun2009gesundheit}, LiveJournal \cite{backstrom2006group}. In work that aligns more closely to our focus on social messaging, Aral et al. \cite{aral2009distinguishing} have looked into the effects of peer-influence and homophily during the diffusion processes. Our work has a more comprehensive scope than \cite{aral2009distinguishing} by integrating both individual-level and group-level features into our cascade prediction model.  

The membership cascade process in the context of social messaging groups differs from previous work in two folds. First, in terms of cascade size, in contrast to previous findings on the structural virality in global-scale cascades \cite{anderson2015global,goel2013structural,watts2002simple}, we observe a relatively smaller scale of cascade in message groups and find a siginificant fraction of short-term group memerbship cascades terminate at the depth of 2 or 3. Such difference is caused by the social messaging group nature of maintaining a compact community size than going ``viral'' \cite{weng2013virality}. Second, due to mechanics specific to application, the cascade process we consider here is locally dependent, with unidirectional contagion dominated mostly by the existing group members. In WeChat, all the groups are by default only visible to group members and grow in a invitation-only fashion. Only the one-hop neighbors of current group members can be invited to the group chat. And this is very dissimilar to the diffusion models in previous work by Backstrom et al.~\cite{backstrom2006group}  which assumes that users can make their own decision to join based on the group influence, even if they are not friends with any of the current group members.



\section{Data}
\label{sec:data}

\noindent {\bf Preliminaries.} Before describing the details of the dataset, we first give a brief overview about WeChat's {\em Group Chat} feature that is central to our study here. While WeChat supports many other important features including {\em Moments} for photo sharing, {\em Friend Radar} for searching nearby friends and {\em Sticker Gallery}, it is important to note that those are beyond the scope of our research focus in this paper.  

On WeChat, each user keeps a brief profile, including demographical information (e.g., gender, age and region)  and address book which saves the contact list of user's friends. We use the tuple $(u, v, T)$ to denote a friend relationship record if user $u$ becomes friend with user $v$ at timestamp $T$. 

\begin{figure}[t]
\centering
\mbox{
	\hspace{-0.1in}
	\subfigure[WeChat group membership]{
		\label{fig:groupchat_ui}
		\includegraphics[width=.48\columnwidth]{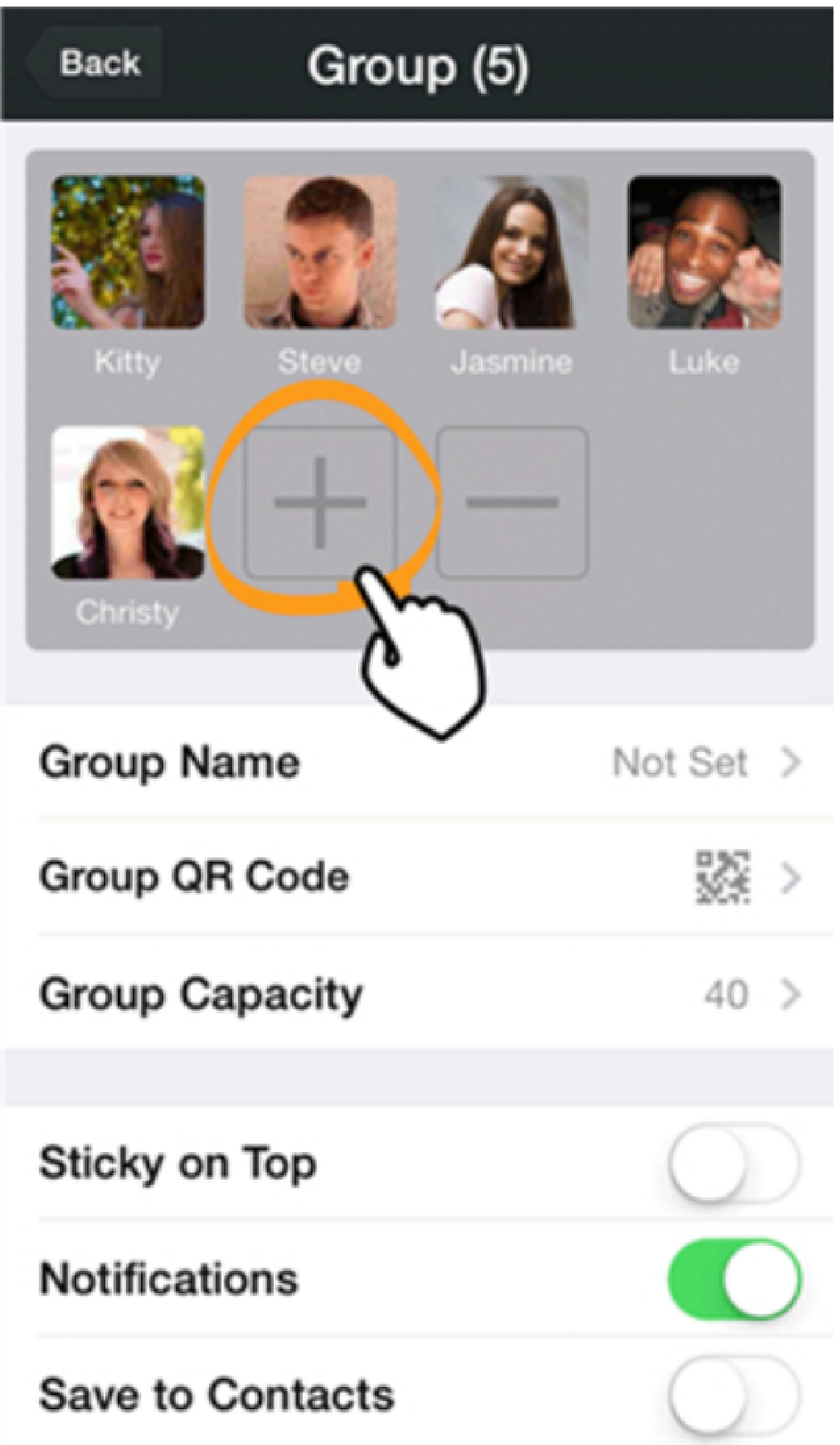}
	}
	\subfigure[Membership invitation]{
		\label{fig:invite_ui}
		\includegraphics[width=0.475\columnwidth]{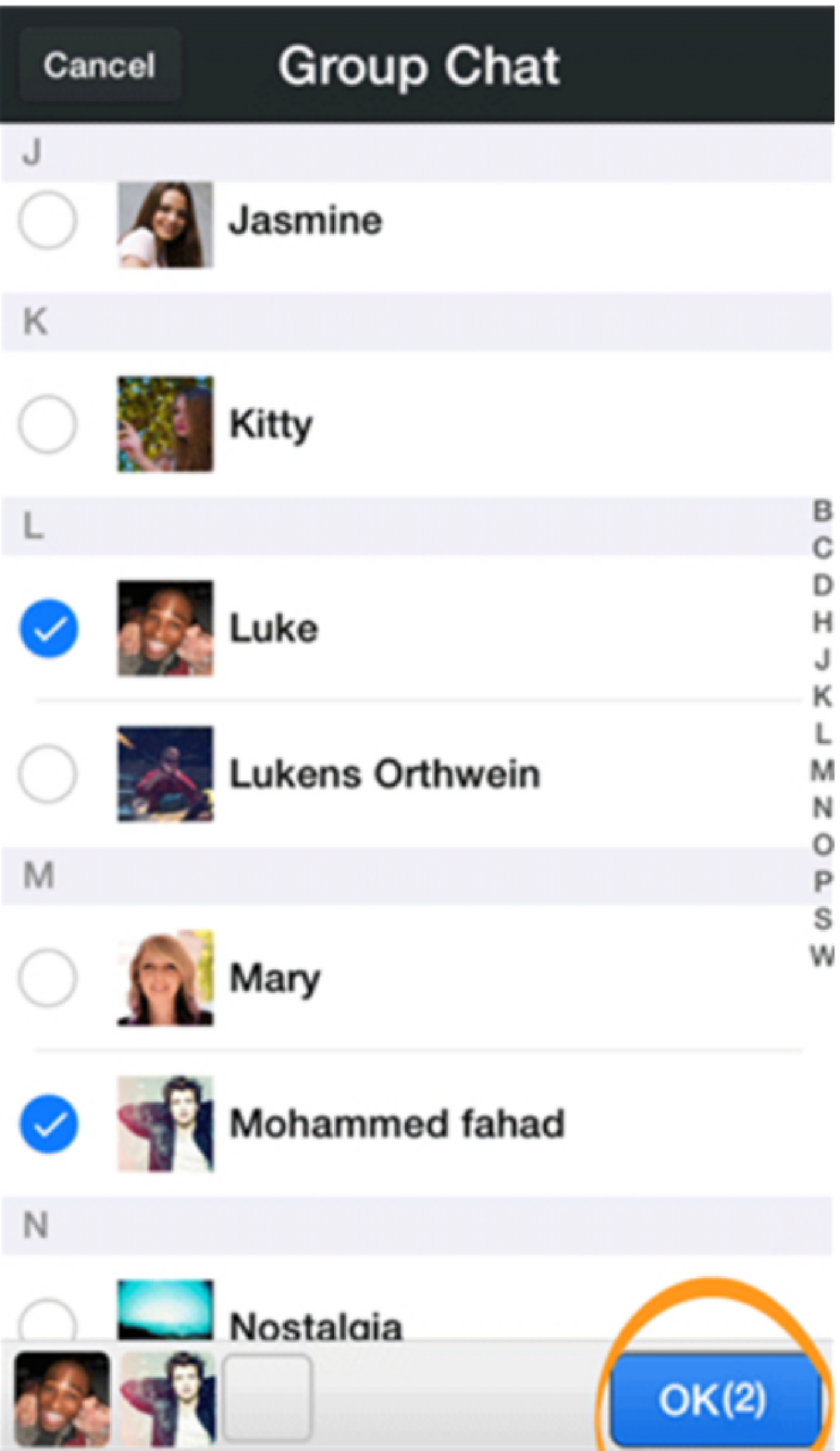}
	}
}
\vspace{-0.15in}
\caption{The WeChat user interaface (UI) of inviting friends to group chat. (a) Group membership UI displays the current users in the group, along with attributes and basic settings for the group such as group name, group capacity etc. Group members can tap the ``+'' button to invite friends into the group chat. (b) The UI of inviting friends to the group chat. Users can browse and select contacts to add, and click ``OK'' to send invitations. When group size is under the capacity 40, invited users will be automatically added into the current group chat without requiring further confirmation. However, under the circumstances that group size exceeds the capacity limit, invited users will have to manually click through the invitation message in order to join the group. The largest WeChat group can have as many as 500 members by default. Sourced from WeChat official feature site \cite{official}. } 
\vspace{-0.15in}
\label{fig:wechat_ui}
\end{figure}

A chat group on WeChat can be analogy to a community, where one can chat with several friends all at once.
There are two ways in which a user can involve in a chat group: one can either initiate a new chat group, or get invited by an existing
member of the group. \figref{fig:wechat_ui} illustrates an example of the WeChat user interface of inviting friends to group chat. We consider $(u, v, \mathcal{C}, T)$ as a successful invitation if user $v$ joins group $\mathcal{C}$ invited by user $u$ at timestamp $T$.


After being a member of a chat group, one can send various forms of messages (e.g., text, photo and voice) to the entire group. We use the tuple $(u, \mathcal{C}, T)$ to denote the
a group chat record if user $u$ send a message to group $\mathcal{C}$ at timestamp $T$.

\vpara{Data Collection and Cleaning.} The data for this study comes from anonymized logs of complete WeChat group messaging activities, collected between July 26th, 2015 to August 28, 2015. 
We first collect all the ~2.3 million groups generated on July 26th, 2015, as our group set of interest. 
We preprocess the data by ignoring groups with less then 5 chat logs--- i.e., we only consider groups that are not born to be dead;  and also filtering groups with users that are in list of monthly spam users (MSU) or monthly inactive users (MIU). The list is maintained and updated by WeChat on a monthly basis. All the initial groups in consideration consist of at least three members. 

\vpara{Data Description.} 
After preprocessing the initial group set, we are left with 474,726 groups for further analysis. We then collect four datasets of interest listed below. Tabel \ref{tbl:data} summarizes statistics of the dataset used for this study.

\begin{itemize}
\item {\bf Group Activity Records $\mathcal{G}$}: It consists of all the temporal group activity records $(u,\mathcal{C},T)$ for each of the sampled group, with $T$ running between July 26th, 2015 to August 28, 2015.  

\item {\bf User Set $\mathcal{U}$}: It consists of all the members belonging to the sampled groups as well as their one-hop neighbors, as of August 28, 2015. Note that we further remove users in the list of MSU or MIU from the user set.

\item {\bf Invitation Records $\mathcal{I}$}: It consists of tuples $(u,v,\mathcal{C},T)$ where user $u$ successfully invites $v$ to join group $\mathcal{C}$ at timestamp $T$ during our data collection period.

\item {\bf Friendship Records $\mathcal{F}$}: It consists of all the tuples $(u,v,T)$ where $u$ and $v$~($u, v\in \mathcal{U}$) become friends with each other at time $T$.
The friend relationships in WeChat are undirected, and we have both $(u, v, T)\in \mathcal{F}$ and $(v, u, T) \in \mathcal{F}$. 

\end{itemize}

\begin{table}[htbp]
\vspace{-0.15in}
\centering
\caption{Summary of data set.}
\label{tbl:data}
\begin{tabular}[htbp]{c|r|r}
\hline \hline
Category & Type 		& Number  \\ \hline
\multirow{3}{*}{Group}	& Total			& 474,726			\\  
	    & Min group size				&3		\\ 
	    & Max group size				&500		\\ \hline
User    & Total		&   245,352,140\\\hline 
Invitation & Total						&	2,013,351		\\ \hline
Friendship & Total						&   624,529,005	\\ \hline
\hline
\end{tabular}
\vspace{-0.1in}
\end{table}

\section{Group Lifecycle Dichotomy}
\label{sec:dichotomy}

One question that we brought up previously is how social messaging groups grow and evolve over time --- their lifecycles and their structural dynamics. As a high-level characteristic, social messaging groups can have a relatively shorter lifespan --- ranging from several hours to months --- as opposed to those online groups seen in social networking sites such as Reddit \cite{buntain2014identifying} and Facebook \cite{park2009being} that can sustain up to years. In this section, we start with discussing the phenomena of lifecycle dichotomy we observe from the group activity temporal data. To do this, we define the lifespan of a social messaging group below.


\begin{definition}
	{\bf Group Lifespan.} We define it by the duration from the timestamp at which a group is initialized, to the timestamp at which no group member sends chat messages anymore.      
\end{definition}

\begin{figure}[htbp]
\centering
\mbox{
	\hspace{-0.1in}
	\subfigure[Histogram]{
		\label{fig:long-short_Hist}
		\includegraphics[width=.5\columnwidth]{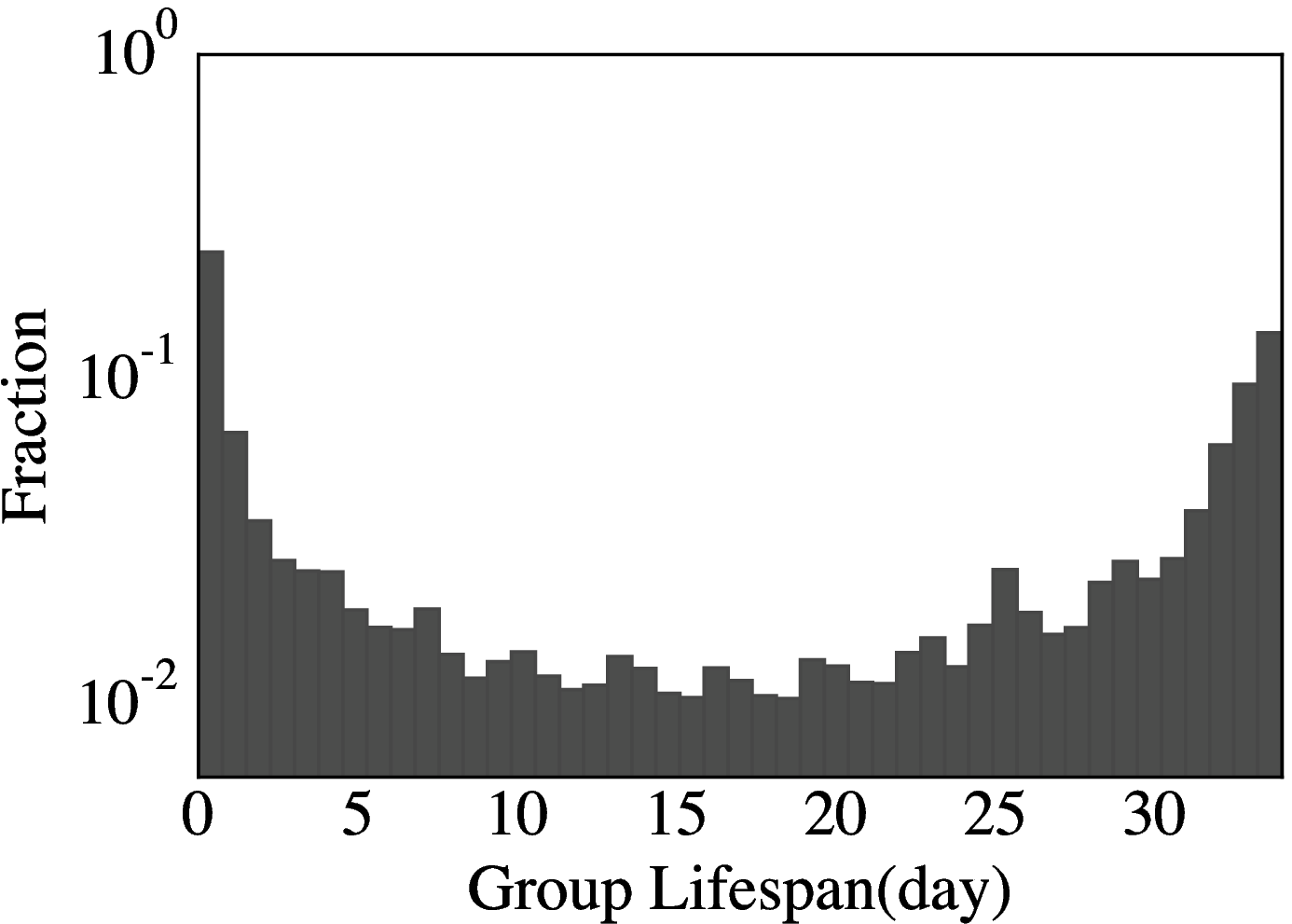}
	}
	\hspace{-0.15in}
	\subfigure[CDF]{
		\label{fig:long-short_CDF}
		\includegraphics[width=.5\columnwidth]{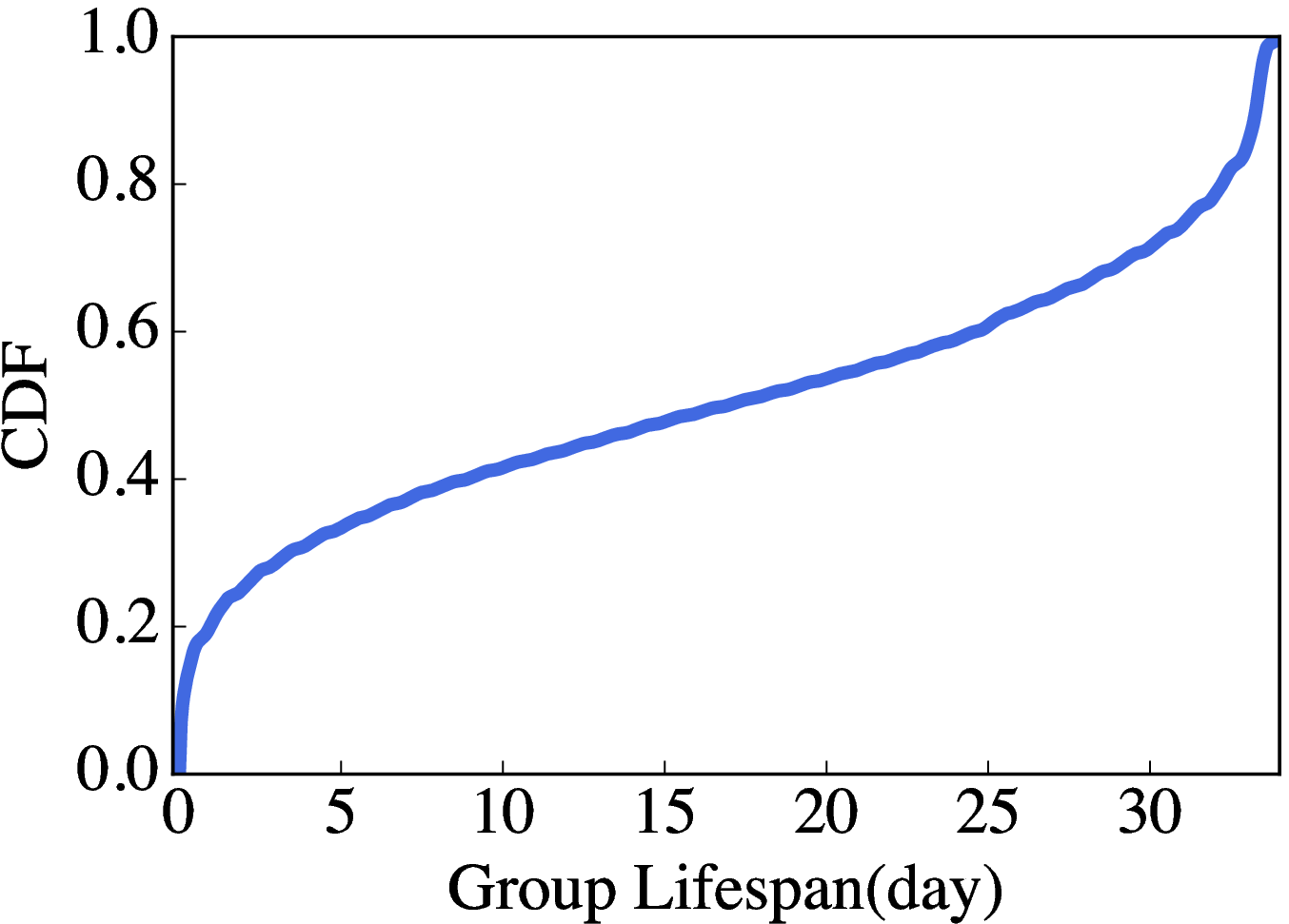}
	}
}
\vspace{-0.15in}
\caption{Group Lifecycle Dichotomy. Left: Histogram of \textbf{\textit{group lifespan}}~(measured by day);
Right: Cumulative distribution function~(CDF) of \textbf{\textit{group lifespan}}.}
\vspace{-0.15in}
\end{figure}

We begin by analyzing the distribution of lifespan among all the 474,726 group samples. Since we stop our data collection on the day of August 28, 2015, the longest lifespan a group can have is 34 days during the period of our observation. \figref{fig:long-short_Hist} and \figref{fig:long-short_CDF} display the distribution and Cumulative distribution function~(CDF) of group lifespan
respectively. A salient observation drawn from the result is that the histogram of group lifespan is dominated by two peaks: one appears on the leftest (near a few hours) and another appears on the rightest side (near one month). This implies a strong dichotomy exists among groups in terms of their lifecycle, and we accordingly make a conceptual difference between two types of groups:

\begin{itemize}
	\item {\bf Short-term groups}: this type of groups emerge and die very quickly, and usually have lifespan ranging from hours to a few days. For example, \figref{fig:long-short_CDF} shows that almost 40\% of groups stop interaction within only a week.
	\item {\bf Long-term groups}: this type of groups can survive a much longer period of time than short-term groups. \figref{fig:long-short_CDF} shows that about 30\% groups fall into this category and can sustain longer than 30 days.
\end{itemize}


The phenomena of lifecycle dichotomy also leads us to the question of how does the lifecycle and growth pattern of social messaging groups correlate with the social functions it is serving? To address this, we manually examine 100 randomly selected groups, among which 60 are long-term groups and 40 are short-term groups, respectively. We categorize these groups according to their social functions~(the title of groups) by hand, and list the details in Table \ref{tbl:case}. Quite interestingly, we find that most short-term groups are event-driven (e.g., travel groups, meeting groups and dining groups), while long-term groups are more relationship-driven (e.g., family groups, colleague groups and friend groups).  

\begin{table}[htbp]
\vspace{-0.15in}
\centering
\caption{Case study by group displayed name.}
\label{tbl:case}
\begin{tabular}[htbp]{c|c|c|l}
\hline \hline
Category & Long 		& Short  & Example\\ \hline 
Travel	    & 0 &8		& Discuss on a short trip\\ 
Meeting    & 1 &   2 & Schedule an official meeting\\ 
Event		&    4 & 13  & Plan a wedding\\ 
Entertain &  5 & 13	&  	Dine together \\ 
Organization & 9						&   0  & Departments of company\\ 
Class & 12						&	4		& Course for GRE test \\ 
Friend & 13 & 0 & Childhood friend\\
Family & 16			& 0			& A family of three\\ 
\hline
\hline
\end{tabular}
\vspace{-0.1in}
\end{table}



\subsection{Group Structure Dynamics}
\label{subsec:structure}

\begin{figure*}[htbp]
\centering
\mbox{
	\hspace{-0.1in}
	\subfigure[Example]{
		\label{fig:structure_example}
		\includegraphics[width=.37\columnwidth]{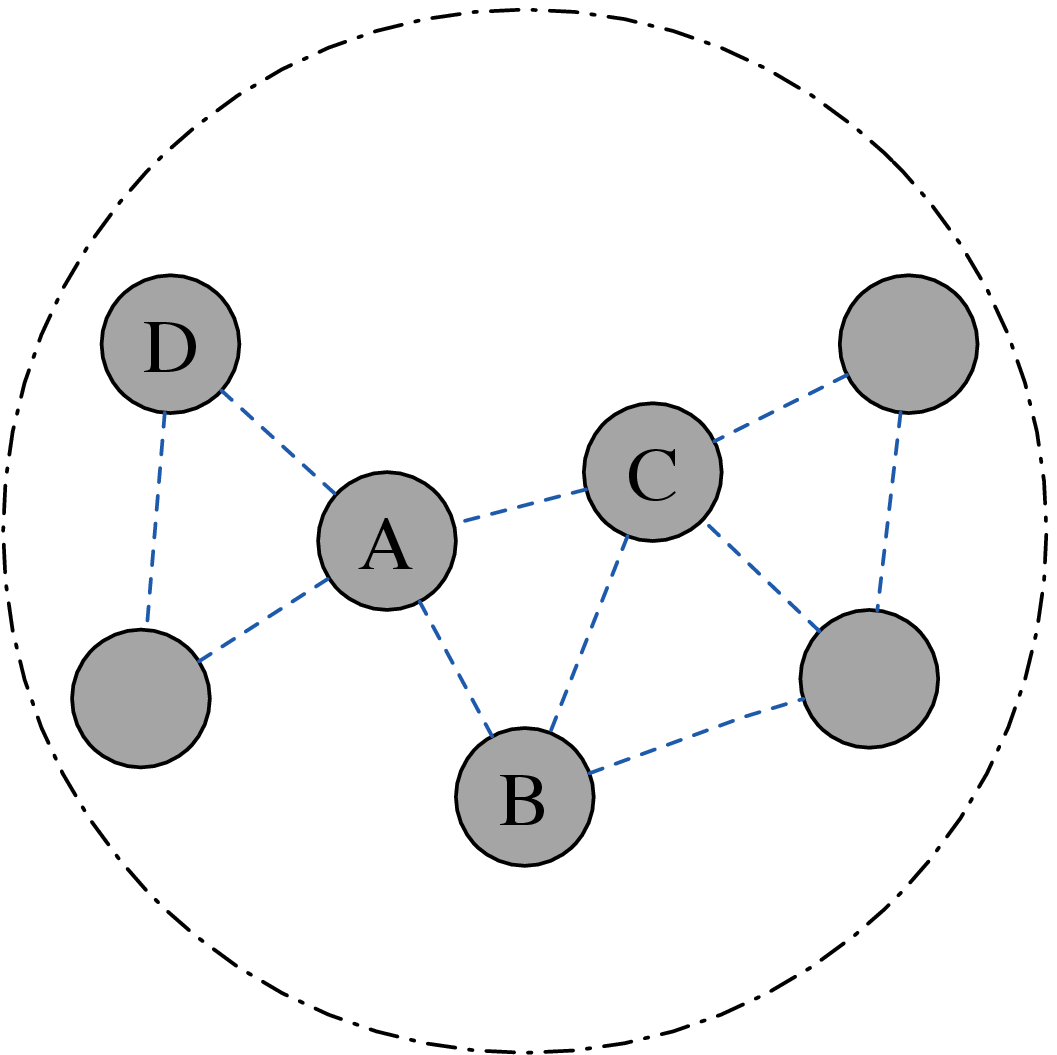}
	}
	\subfigure[Open triads]{
		\label{fig:open_triads}
		\includegraphics[width=.54\columnwidth]{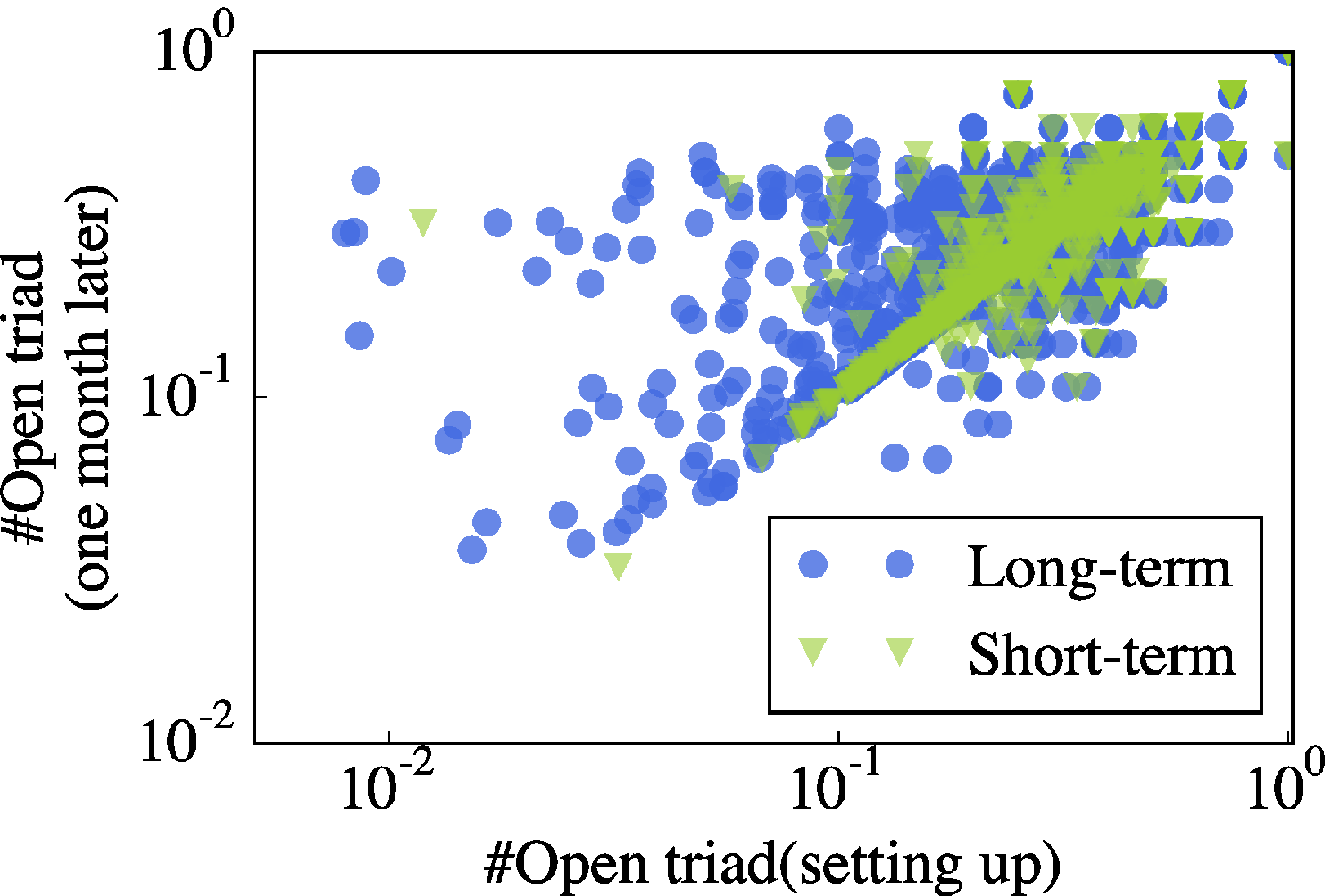}
	}
	\subfigure[Closed triads]{
		\label{fig:closed_triads}
		\includegraphics[width=.54\columnwidth]{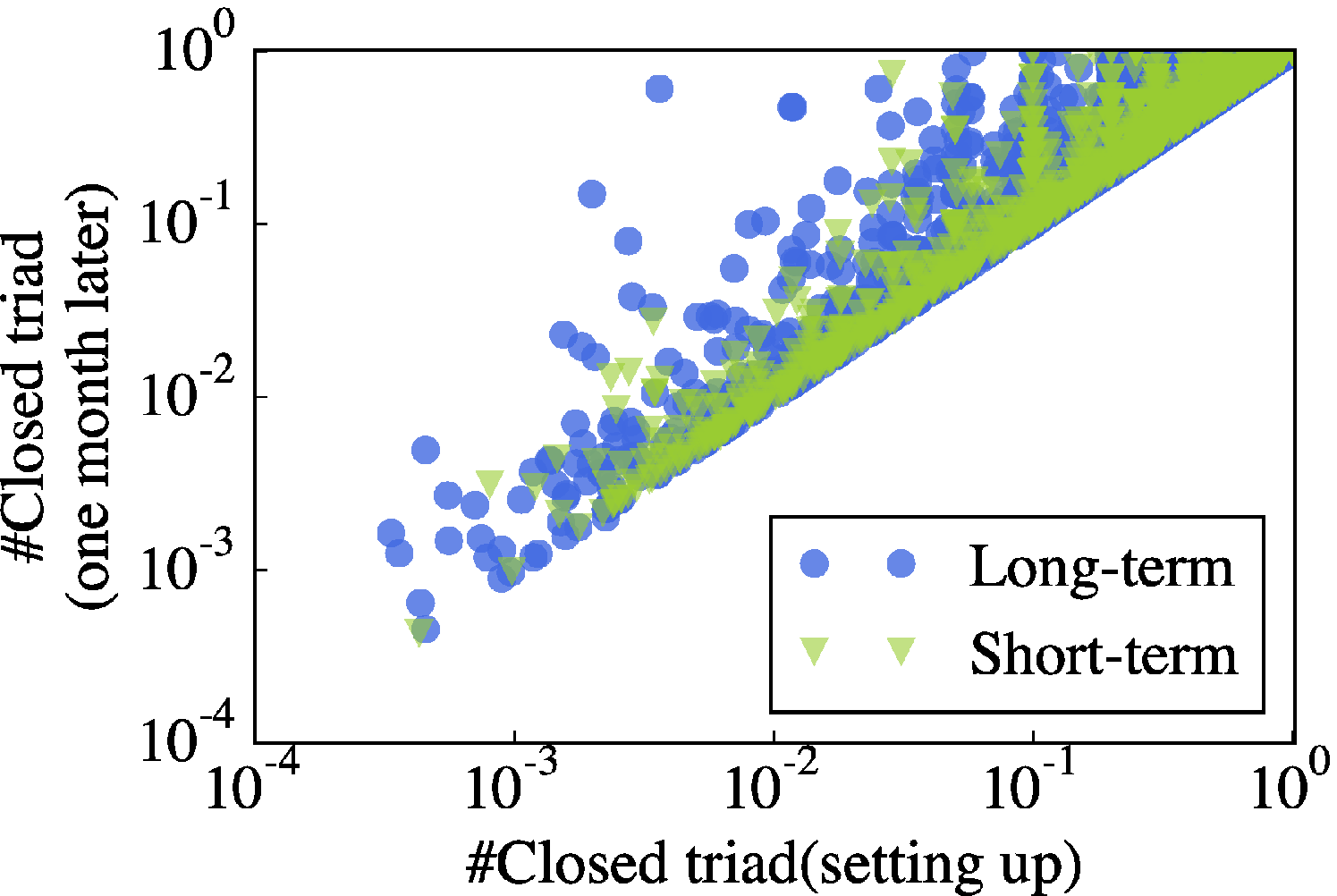}
	}
	\subfigure[Edge density]{
		\label{fig:edge_density}
		\includegraphics[width=.54\columnwidth]{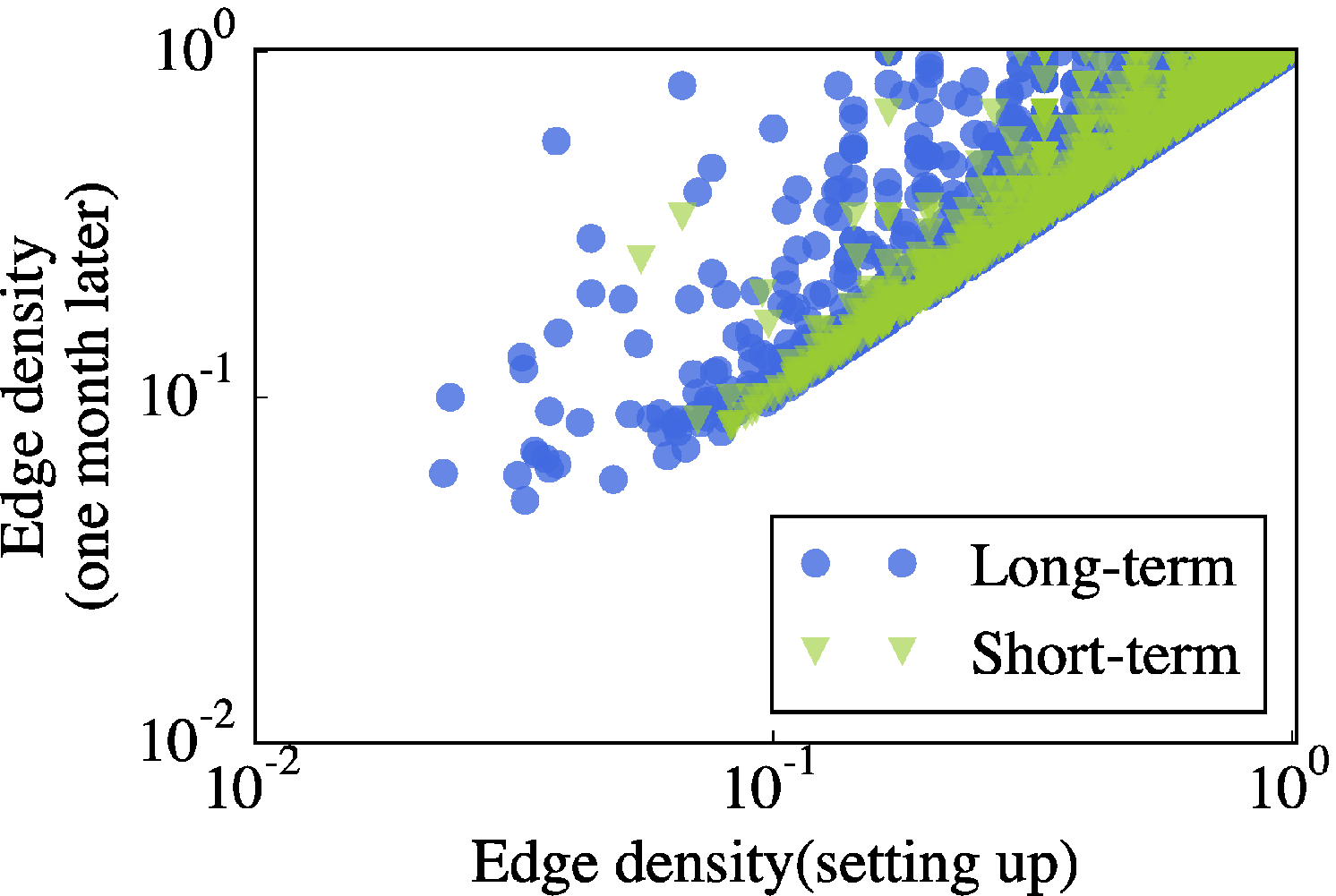}
	}
}
\vspace{-0.15in}
\caption{Group Structure Pattern. (a): An example of group friendship networks. Group members and friend relationships are represented as nodes and dot lines, respectively.
	For this example, we can see $A$, $B$ and $C$ are connected as a closed triad; $A$, $C$ and $D$ are connected as an open triad. The edge density of this group is 0.476.
	(b)~(c): Horizontal axis is the normalized number of open/closed triads at the setting up of a WeChat group,
	and vertical axis is the normalized number of open/closed one month later. (c):  Horizontal axis is the edge density at the setting up of a WeChat group, and veritcal axis
is the edge density one month later.} 
\vspace{-0.15in}
\end{figure*}

In this subsection, we move on to study the underlying structural change of messaging groups over time.
We investigate several representative structural features (e.g., open triad count, closed triad count and edge density), and
quantitatively analyze the how these features evolve in a different pattern with respect to the long-term and short-term groups, respectively. 


\vpara{Triad Count.}
The studies about transitivity in social networks~\cite{holland1971transitivity}
suggest that the local structure in social networks can be expressed
by the {\em triad count}. In WeChat groups, we try to examine whether long-term and short-term groups
show different transitivity patterns. We take into account both the open triad count and close triad count, based on the friendship networks structure of sampled WeChat groups. To illustrate this, \figref{fig:structure_example} shows an example of a small WeChat group friendship networks, in which nodes $A$, $B$ and $C$ form a closed triad; nodes $A$, $C$ and $D$ is considered  an open triad. 

\vpara{Edge Density.}
We also consider the feature of internal {\em edge density} of a group, which is defined by the fraction of edges (friendships) within the group among all the possible edges when the group is fully connected.   

\vspace{0.5em}

To see how these structural features change over time, we take two snapshots for the groups: one at the time when the groups are initialized (we choose $\sim$10 minutes in this study), another after one month being setting up. We consider long-term and short-term groups separately in order to see the different patterns of structural patterns between these two. We also remark here that although short-term groups may stop messaging interactions at some point, members within the groups are still likely to build friendship as long as they maintain the group membership, and thus affect the underlying friendship network structure for potentially longer period of time. 

\figref{fig:open_triads}, \figref{fig:closed_triads} and \figref{fig:edge_density} show the results for feature dynamics of open triad count, close triad count and edge density, respectively. 
Note that if the structure of groups are not changing at all, we would expect to see a scatter plot centering around the diagonal line of $y=x$ (with normalization). 
From the visualization results, it is interesting to first observe the different evolution patterns exhibited between short-term groups and long-term groups --- the long-term ones show stronger dynamics in terms of the underlying friendship structure features while most short-term groups are less likely to develop friendship over time.

We infer such dichotomy in structure dynamics is related to the social roles and functions for the social groups to be setup. For example, a colleague's group served for long-term communications is more likely to develop social connections between members, as opposed to a group setup for some specific social event.  

\subsection{Cascade Tree Pattern}
\label{subsec:cascade}
Beside studying the friendship structure, we also discuss the group formation processes, namely by investigating the group membership invitation cascading tree structure. We start with defining the {\em group cascade tree} below.  

\begin{definition}
	{\bf Group Cascade Tree.} A directed graph where each group member is a node, and a directed edge from $u$ to $v$ is constructed if $u$ (inviter) successfully invites $v$  (invitee) to the group. The tree is rooted at the user who initiated the group. Cycles are impossible since inviters always join the group earlier than invitees. 
\end{definition}

\begin{figure}[htbp]
\centering
\mbox{
	\hspace{-0.1in}
	\subfigure[Example of long-term group]{
		\label{fig:longexample}
		\includegraphics[width=.5\columnwidth]{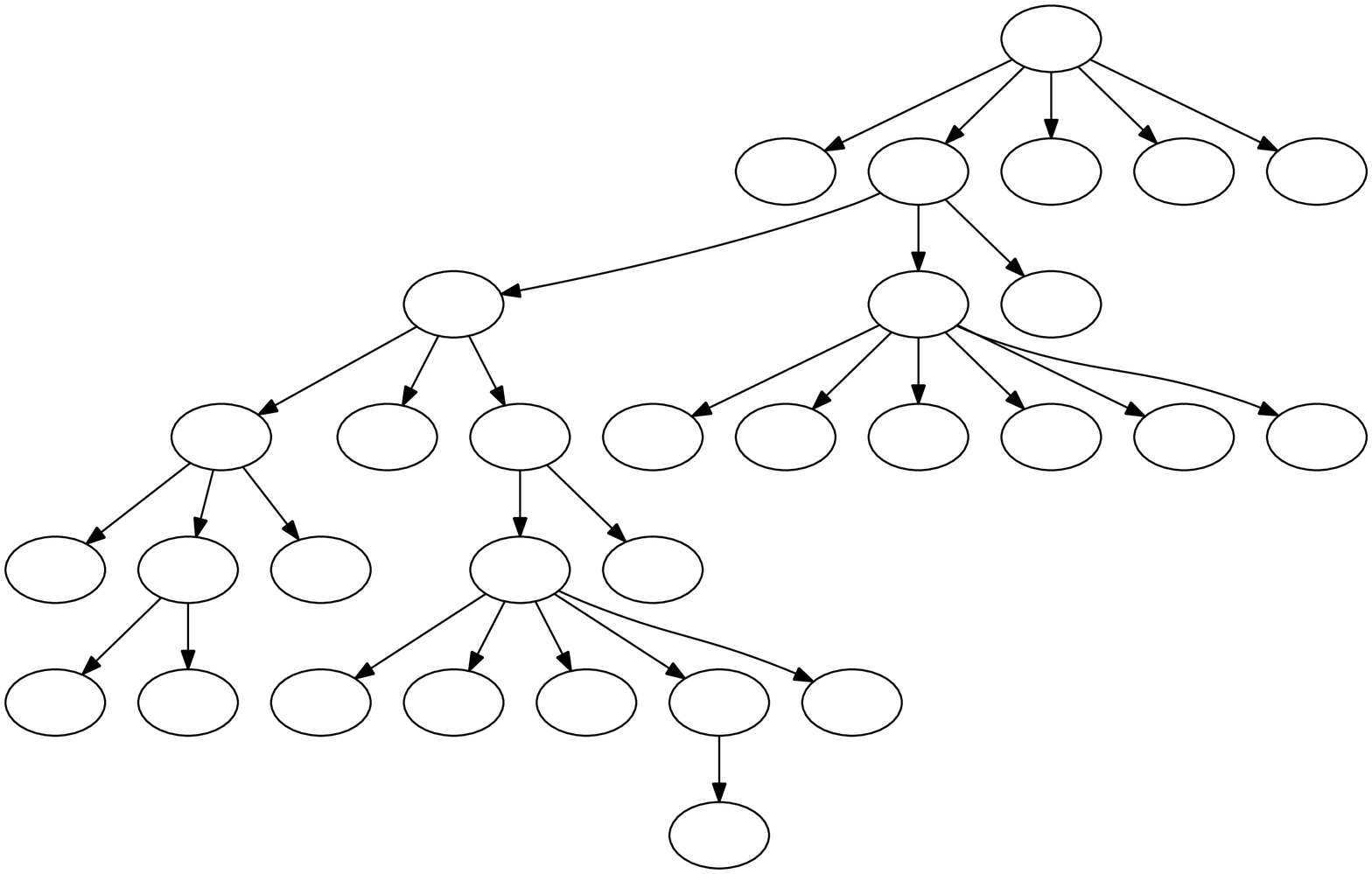}
	}
	\subfigure[Example of short-term group]{
		\label{fig:shortexample}
		\includegraphics[width=.5\columnwidth]{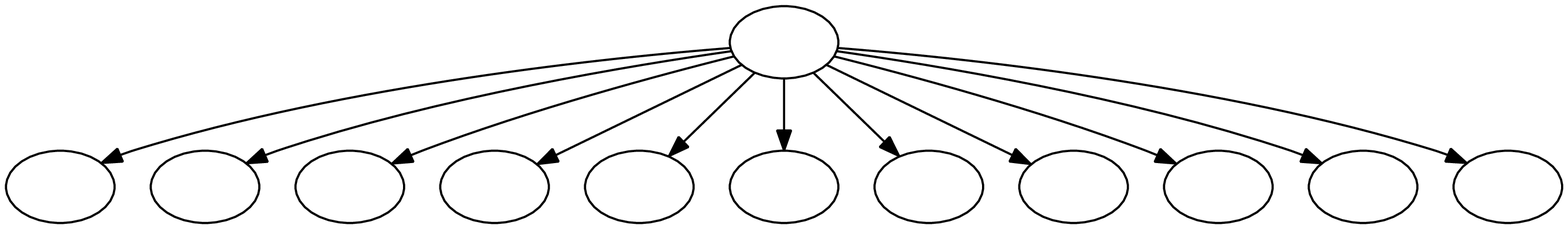}
	}
}
\vspace{-0.15in}
\caption{Example of WeChat group cascade tree for long-term group and short-term group, respectively. }
\end{figure}

\begin{figure*}[htbp]
\centering
\mbox{
	\hspace{-0.1in}
	\subfigure[Group size]{
		\label{fig:group_size}
		\includegraphics[width=.5\columnwidth]{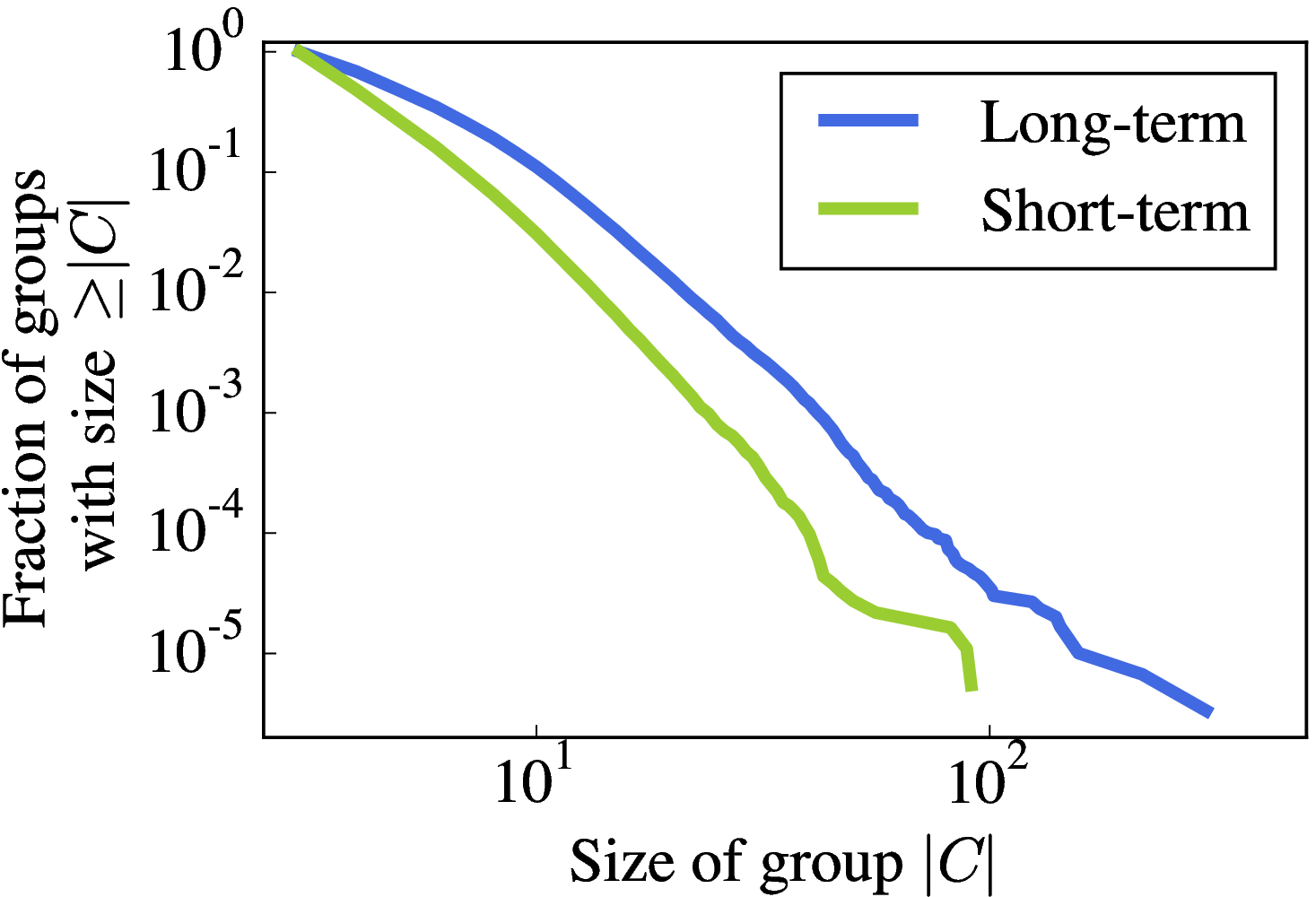}
	}
	\subfigure[Subtree size]{
		\label{fig:subtree_size}
		\includegraphics[width=.5\columnwidth]{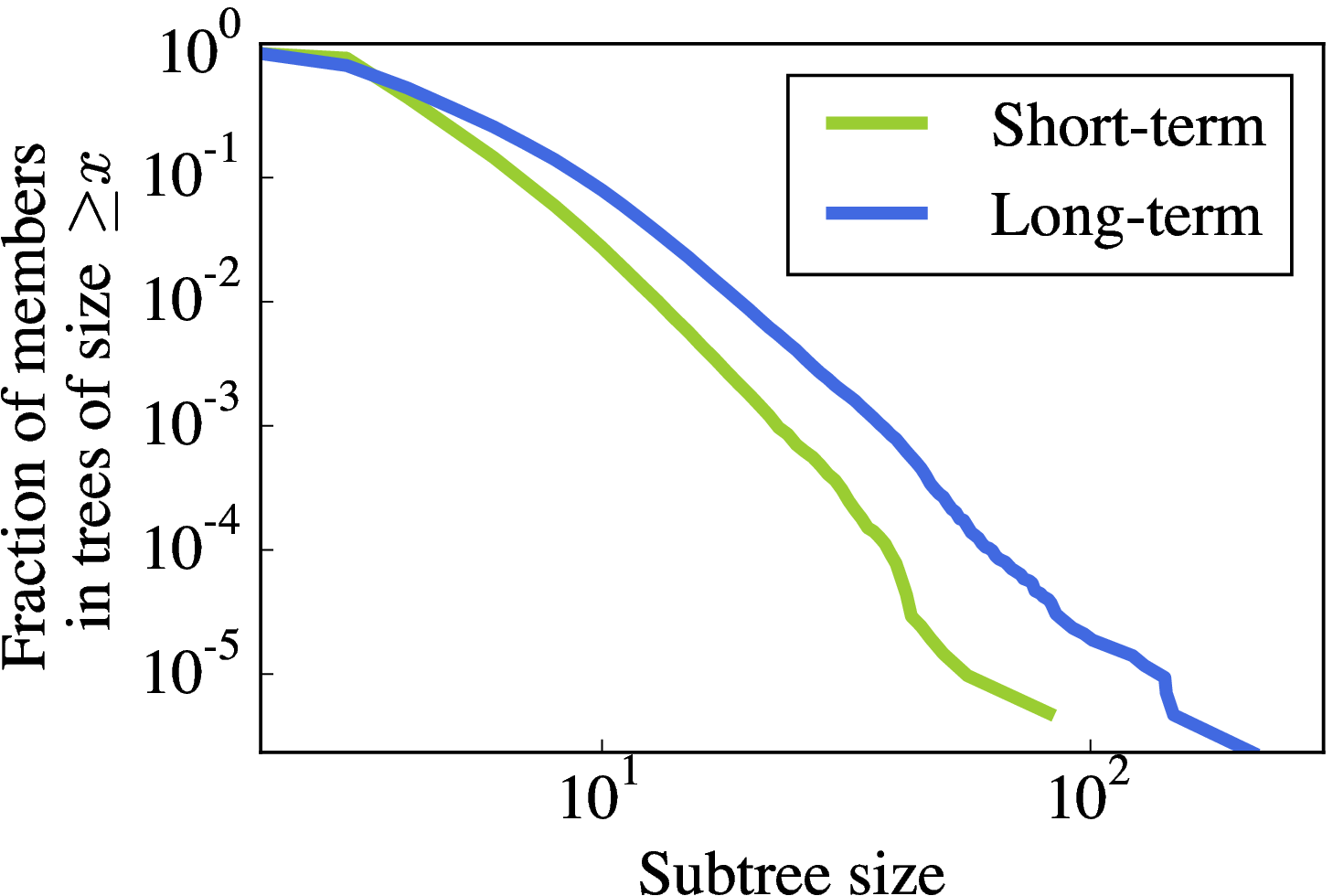}
	}
	\subfigure[Depth]{
		\label{fig:depth}
		\includegraphics[width=.5\columnwidth]{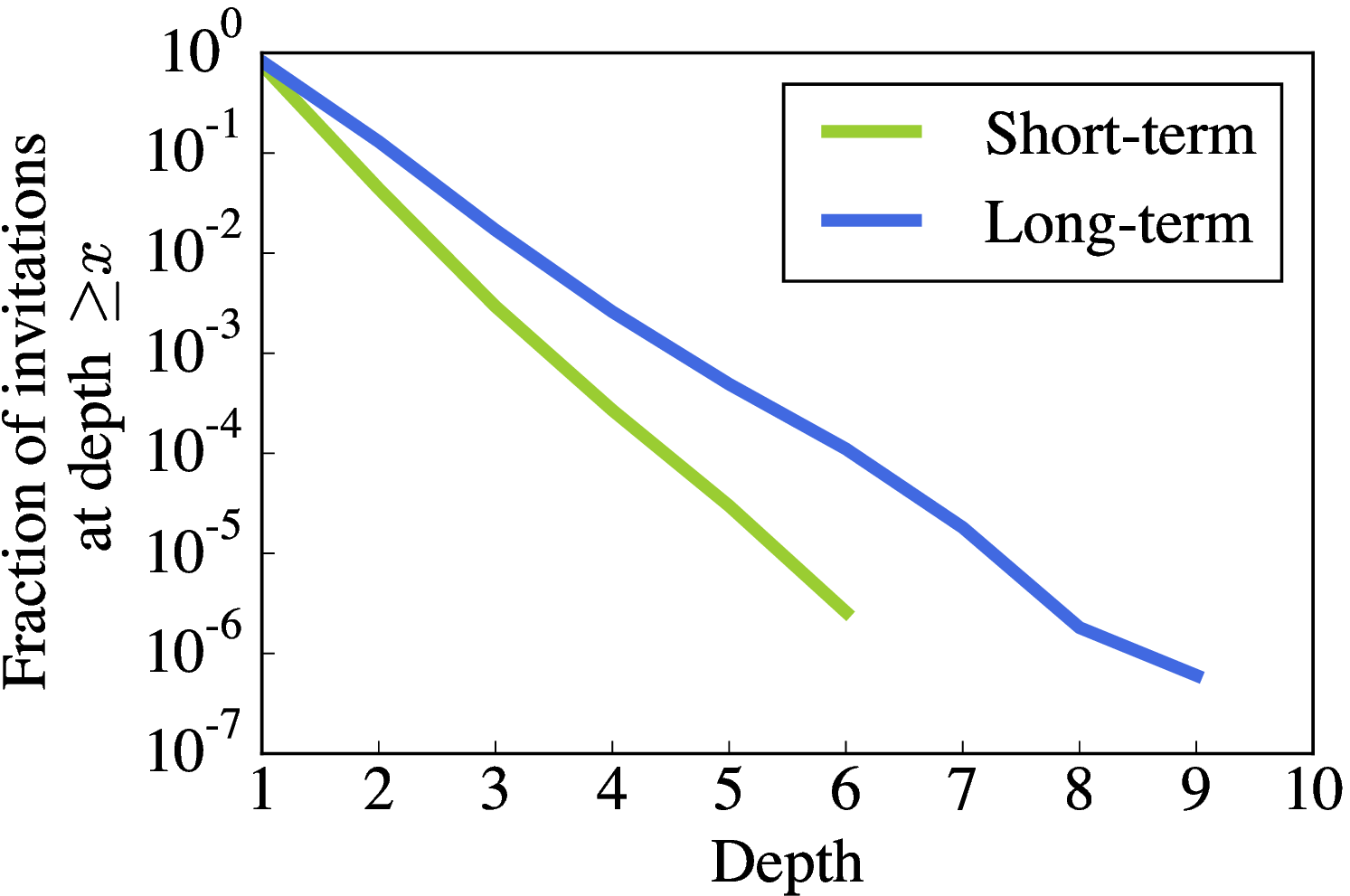}
	}
	\subfigure[Wiener index]{
		\label{fig:windex}
		\includegraphics[width=.5\columnwidth]{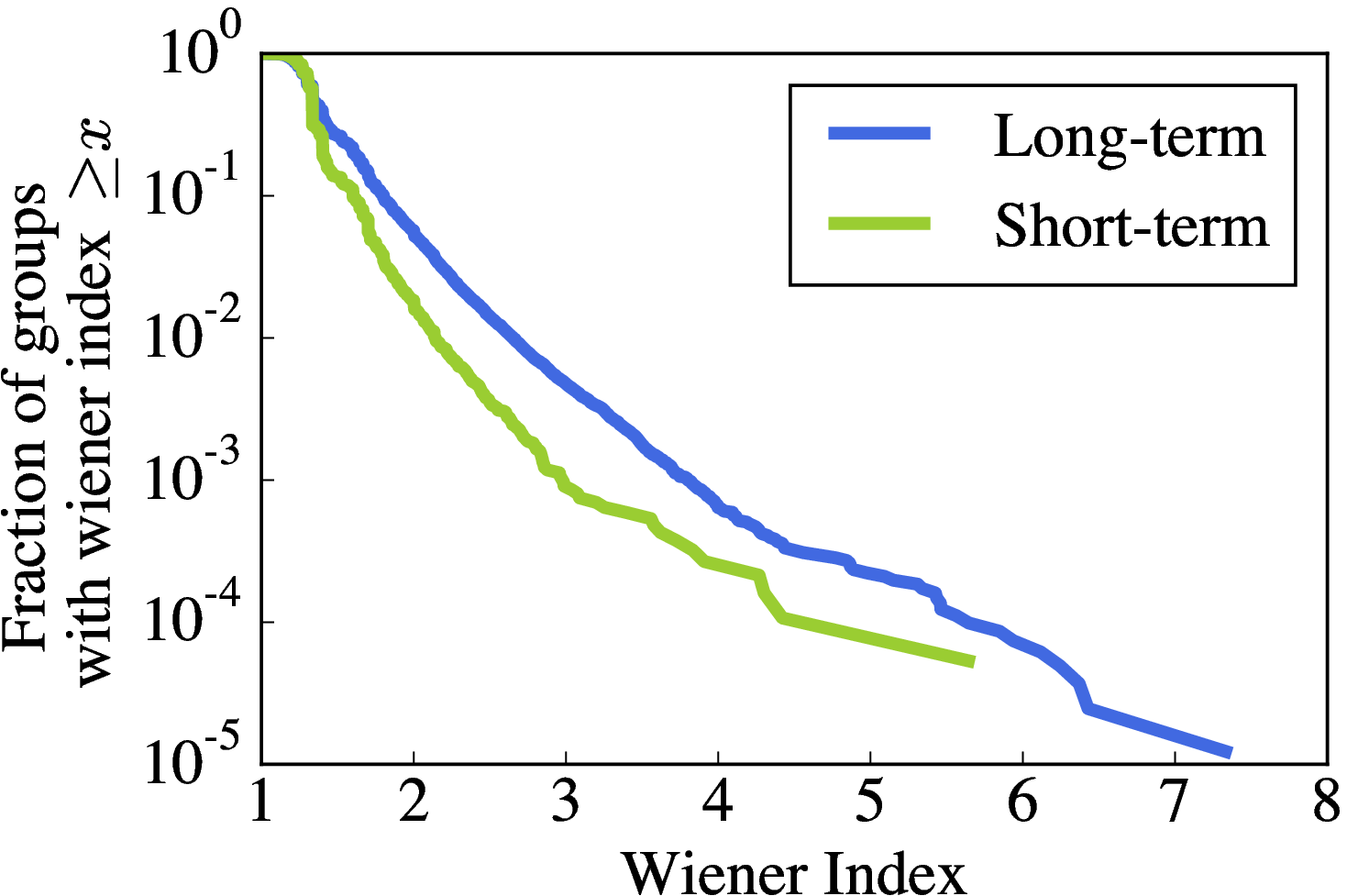}
	}
}
\vspace{-0.15in}
\caption{Cascade tree related feature distributions. (a): Distribution over group size. Vertical axis is the fraction of groups with size larger than $\left|C\right|$.
(b): Distribution over subtree size. Vertical axis is the fraction of non-singleton members in trees of specific size.
(c):  Distribution over cascade depth. 
(d): Distribution over Wiener index.} 
\vspace{-0.1in}
\end{figure*}

To show how long-term and short-term groups differ in terms of cascade tree structure, \figref{fig:longexample} and \figref{fig:shortexample} show the examples for two types of WeChat group cascade tree. 
We find that long-term groups tend to exhibit a deeper tree structre with more branchings; whereas many short-term group cascade trees display an approximate star graph structure with most members being the leaves of the root node. In order to quantify such difference, we consider here four representative features concerning the structure of cascade trees. 

\vpara{Cascade Size.} We start our analysis on cascade tree by examining the total number of nodes in the cascade tree, i.e., group size.
\figref{fig:group_size} shows the normalized distribution of cascade tree size for both types of groups.
We find that long-term groups tend to have larger size (up to 500 by default) while the size of short-term groups diminishes around 100. This is not very surprising since long-term groups can be advantageous in gaining more members give a longer timespan for growth. 

\vpara{Invitation as a Function of Cascade Depth.} A natural way to measure the
difference in cascade tree between long-term groups and short-groups is to examine the distribution of cascade depth at which the invitation occur. We measure the cascade depth for each of the invitation happens during our observation period, which is defined by the number of steps from the root to the group member in the cascade tree.
\figref{fig:depth} shows the normalized distribution of cascade depth among all the invitations in our dataset.
We observe that more invitations occurs far from the root in long-term groups than that in short-term groups.
For example, 10\% of invitations in long-term groups occur at depth 3 or greater; whereas for short-term groups, only less than 1\% of invitations occur at depth 3 or greater. 

\vpara{Invitation as a Function of Subtree Size.} 
Finally, we measure the difference of cascade tree structure between long-term groups and short-term groups by measuring the size of subtree for each node in the cascade tree.
In \figref{fig:subtree_size}, we show the distribution of the cascade subtree size for each node resides in the cascade tree, aggregated among all the sampled groups. 
Again, we observe substantial difference between long-term groups and
short-term groups. For example, 30\% of nodes in the cascade tree of long-term groups have subtree size greater than 10; whereas only 10\% of nodes have subtree size greater than 10 in short-term groups.

\vpara{Structural Virality.} We also quantify cascade trees by measuring
their \textit{structural virality} as that used in \cite{goel2013structural}. Structural virality, also know as Wiener index, is useful for disambiguating  between shallow,
broadcast-like diffusion and the deep branching structures. Wiener index is defined by 
the average distance between any two nodes in the cascade tree. 
For example, the cascade trees in \figref{fig:longexample} and \figref{fig:shortexample} have  Wiener indexes of 3.99 and 1.83, respectively.
In \figref{fig:windex}, we show the distribution of Wiener index of cascade trees for both long-term and short-term groups. We observe that more than 99\% of short-term groups
have Wiener index smaller than 2, which implies that most membership cascades happen in a broadcast fashion, settled mostly by the root node.

\subsection{Group Lifecycle Prediction}

The strong dichotomy of group lifecycle and structure dynamics leads us to a natural modeling and prediction questions —-- how separable are the long-term and short-term groups by taking into account the structural,  behavioral as well as demographical features? Can we predict whether a social group will grow and persist in the long run by analyzing the structural and behavioral patterns exhibited by the group at its early stage? In this section, we address both issues through analyzing the snapshots of millions of groups, combining a broad range of features. 




\subsubsection{Separability Model}
In this model, we consider the task of predicting whether a group is long-term or short-term, from features including the underlying group network structure, the membership cascade tree properties, and the demographics entropy of group members. The full list of features can be found in Table \ref{tbl:feature}, where we are only using group-level ones for this task. 

To train the separability model, we construct the training dataset by labeling groups with less than 5 days of lifespan as negative examples, and groups with longer than 25 days of lifespan as positive examples. We represent each group as a feature vector extracted one month after groups are built, and further train the dataset using support vector machine~(SVM)~\cite{fan2008liblinear} with 10-fold cross validation. 

\begin{table}[htbp]
\vspace{-0.15in}
\centering
\caption{Feature contribution analysis on the separability of long-term and short-term groups.(\%)}
\label{tbl:separability}
\begin{tabular}{c|c|c|c|c}
\hline
\hline
\textbf{Features used} & \textbf{AUC} & \textbf{Prec.} & \textbf{Rec.} & \textbf{F1} \\
\hline
All Features    &     \textbf{66.62}      &     63.23      &    57.66        &    60.32          \\  
 -Structure    &     64.75 &     59.36       &     62.83       &     61.04       \\ 
-Cascade    &     65.36      &     \textbf{64.49}       &     47.67       &     54.82         \\ 
-Demographics   &     65.24      &     57.35       &     \textbf{65.71}        &     \textbf{61.25}         \\ 
\hline
Random Guess & 50.00 & 50.00 & 50.00 & 50.00 \\
+Structure & \textbf{64.21}  & 61.98 & 42.51 & 50.43 \\
+Cascade & 61.23 & 57.35 & \textbf{65.71} & \textbf{61.25} \\
+Demographics & 62.77 & \textbf{63.18} & 41.41 & 50.03\\
\hline
\hline
\end{tabular}
\vspace{-0.1in}
\end{table}

The prediction results are shown in Table \ref{tbl:separability}. We find that highest classification accuracy (66.62\% AUC) can be obtained with full set of features. We further investigated how each set of features (i.e., structure, cascade and demographics) affects the training performance by considering only one at a time. And we find that the set of structural features by itself can yield high accuracy, which again confirms that strong distinctions exist between short-term and long-term group structures. 


\subsubsection{Early Prediction of Group Lifecycle}

Given the strong separability between the long-term and short-term groups, we pose a fundamental question of how well can we  predict if a group can grow and persist in the long run, from the features exhibited in its early age?

\begin{table}[htp]
\vspace{-0.15in}
\centering
\caption{Group lifecycle early prediction performance results(\%). We train the classifier using all the group-level features.}
\label{tbl:early}
\begin{tabular}{c|c|c|c|c}
\hline
\hline
\textbf{Features used} & \textbf{AUC} & \textbf{Prec.} & \textbf{Rec.} & \textbf{F1} \\
\hline
1 hour	&	57.95					&	54.16	& 56.80		& 55.45 \\
1 day    &		\textbf{65.08}				&	{\bf 61.92}	& {\bf 53.38}		& {\bf 57.34} \\
5 days	 &		65.46				&	62.52	& 54.11     & 58.01 \\
10 days	 &		65.57				&   62.48	& 56.81		& 59.51 \\
20 days	 &		65.76				&   62.78	& 56.56		& 59.51 \\	
1 month    &     66.62      &     63.23     &    57.66       &   60.32         \\  
\hline
\hline
\end{tabular}
\vspace{-0.1in}
\end{table}
The way we implement the early prediction model is largely similar to the separability model previously except for the subtle difference that group features (see group-level features in Table \ref{tbl:feature}) are extracted at earlier timestamps. Specifically, for each group in our training set, we take multiple snapshots at the age of 1 hour, 1 day, 5 days, 10 days, 20 days and 1 month, and calculate the feature vector accordingly. We repeat similar procedure to train the dataset with respect to features extracted at various timestamp, and compare the training performance. Table \ref{tbl:early} shows the prediction performance results at different stages. We find that features extracted one day after the groups being set up can yield AUC accuracy as high as 65.08\%, which is almost as good as the prediction accuracy of 66.62\% when adopting features at timestamp of 1 month.

The results of early prediction model reassure that the likelihood for social messaging group to grow in the future can be well inferred from its very early age (e.g., 1 day). Such predictability is in contrast to previous study on predicting the longevity of online social communities \cite{kairam2012life} which requires features at the age of {\em months} for making short-term prediction and {\em years} for making  long-term prediction. And again this is partly due to the different nature of social messaging groups and online communities in terms of the lifecycle.




\section{MEMBERSHIP CASCADE PROCESS}
\label{sec:cascade}

Now we have modeled the growth and evolution of social messaging groups from a group-level perspective. In this section, we approach the problem with a focus on the individual-level and study the membership cascade process by which group gain new members. 

\begin{figure}[htbp]
\centering
\includegraphics[width=1.\columnwidth]{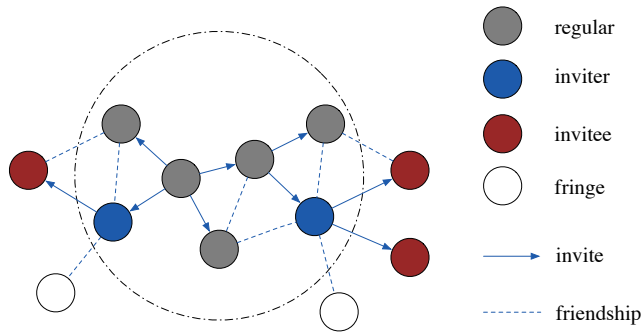}
\vspace{-0.15in}
\caption{Graphical example of WeChat groups' cascade process model. At some timestamp $T$, some members in a group become active (denoted by  blue) and select their friends (denoted by red) to the group chat.}
\label{fig:cascadeExample}
\vspace{-0.1in}
\end{figure}

To start with, we introduce a group membership cascade model, as illustrated in \figref{fig:cascadeExample}. The model captures two important roles: {\em inviter} --- a group member who sends invitation to friend(s), and {\em invitee} --- the individual in the inviter’s ego networks who gets invited to the group chat\footnote{On WeChat, instead of sending group invitation to any registered user, one can only invite his/her current friends into the group chat.}. For instance, the big dotted circle in \figref{fig:cascadeExample} encompasses all the current members within a group. There are two essential steps behind each invitation: 1) a member in a group become active (denoted by blue in \figref{fig:cascadeExample}), and 2) the active member selects his/her friends (denoted by red in \figref{fig:cascadeExample}) into the group chat.   


\subsection{Membership Cascade Pattern}

\subsubsection{Behavioral Pattern}

To have a better understanding of membership cascade pattern, it is important to first study group members' behavioral pattern. For example, an interesting question would be how often do people invite their friends into the group chat once they become a group member? This can also be phrased as how often do membership cascade happen in social messaging groups? In this subsection, we provide some empirical findings concerning members' invitation behavior pattern measured by the concepts of {\em invitation interval} and {\em first invitation latency} defined below.

\begin{definition}
	{\bf Invitation Interval} is defined as the time interval between any two consecutive
invitations from a group member. Additionally, {\bf First Invitation Latency} is defined as the interval
between the timestamp at which a user joins a group (invited by some existing member) and the timestamp when he/she, for the first time, invites another friend to the same group.
\end{definition}

Intuitively, investigating the action of a group member's first invitation is useful since it signifies  how well has he/she been adapting to the current group, and how strong the sense of relevance he/she has with respect to the current group.   
\begin{figure}[htbp]
\centering
\mbox{
	\hspace{-0.1in}
	\subfigure[First invitation latency]{
		\label{fig:member_invitation}
		\includegraphics[width=.5\columnwidth]{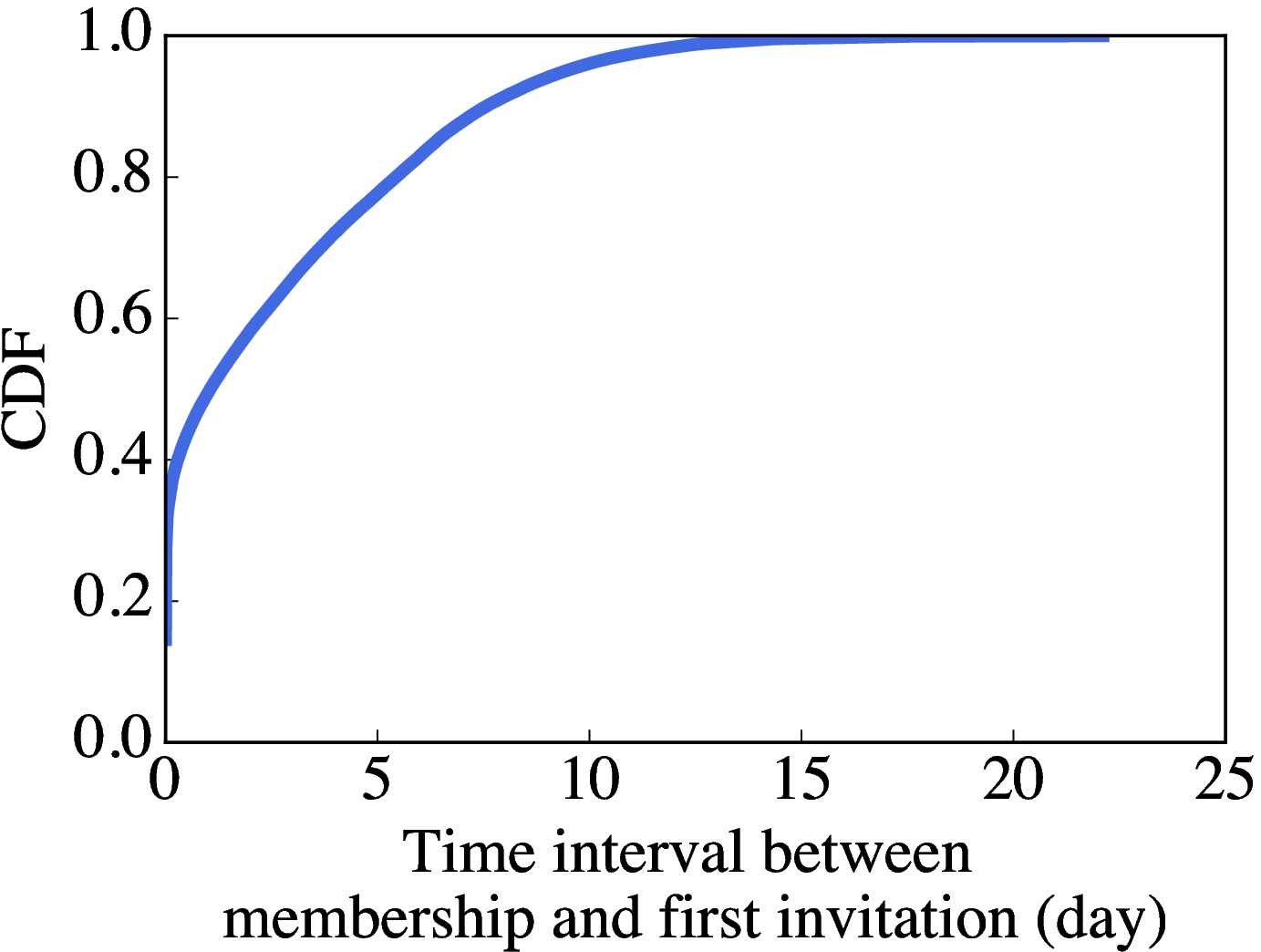}
	}
	\hspace{-0.15in}
	\subfigure[Invitation interval]{
		\label{fig:invitation_invitation}
		\includegraphics[width=.5\columnwidth]{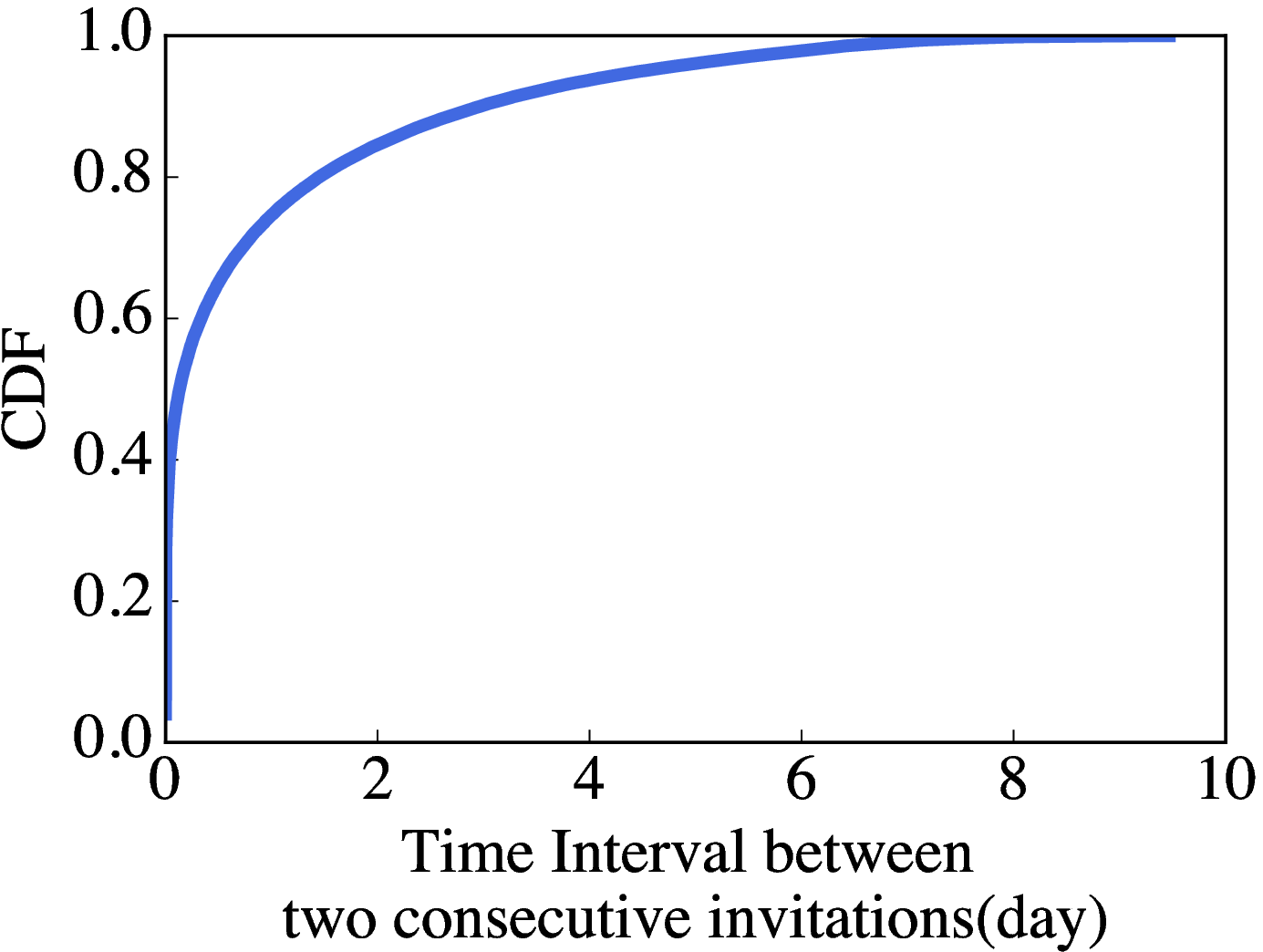}
	}
}
\vspace{-0.15in}
\caption{Dynamic pattern of invitations. Left: Cumulative distribution function~(CDF) of \textbf{\textit{first invitation latency}}~(measured by day);
Right: Cumulative distribution function~(CDF) of \textbf{\textit{invitation interval}}~(measured by day). The invitations in WeChat group display a highly time-sensitive pattern.}
\vspace{-0.10in}
\end{figure}

To address above questions, we obtain the distribution of invitation interval and first invitation latency, aggregating over each member in each group. \figref{fig:member_invitation} and \figref{fig:invitation_invitation} show the CDF curve of first invitation latency
and invitation interval, respectively. We observe the invitations in WeChat groups are highly time-sensitive. On the one hand, when one is invited to a group, he/she tends to start invite other people soon.
For example, about 80\% of the first invitations happen within 5 days after the inviter joining the group.
On the other hand, we find that members suffer from a longer latency in sending their first invitations than the invitation interval in general. For example, more than 80\% of consecutive invitations happen within 2 days of interval.

\subsubsection{The Influence of Local Structure}

\begin{figure*}[p]
\centering
\mbox{
	\subfigure[Example of local structure]{
		\label{fig:ego}
		\includegraphics[width=.45\columnwidth]{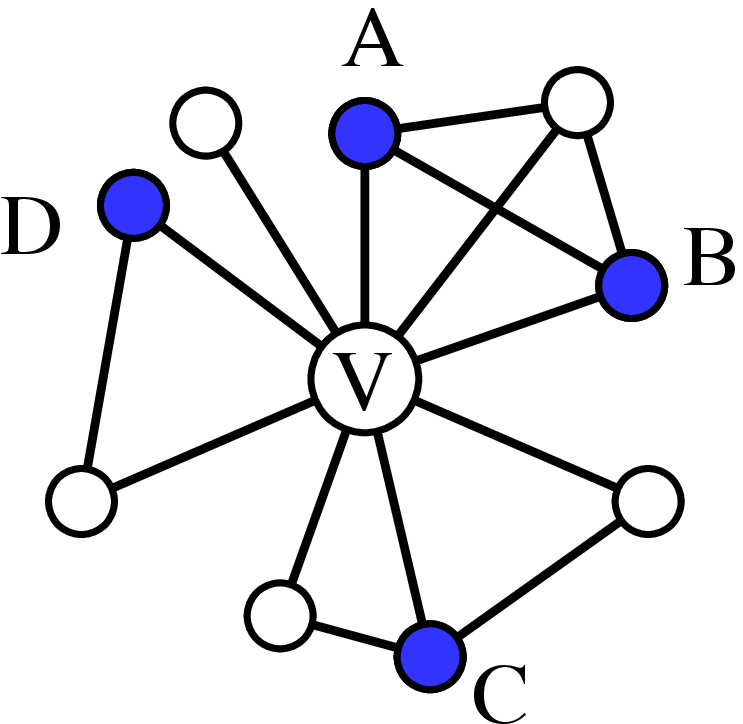}
	}
	\subfigure[\#Friend in group] {
		\label{fig:kfriend}
		\includegraphics[width=.78\columnwidth]{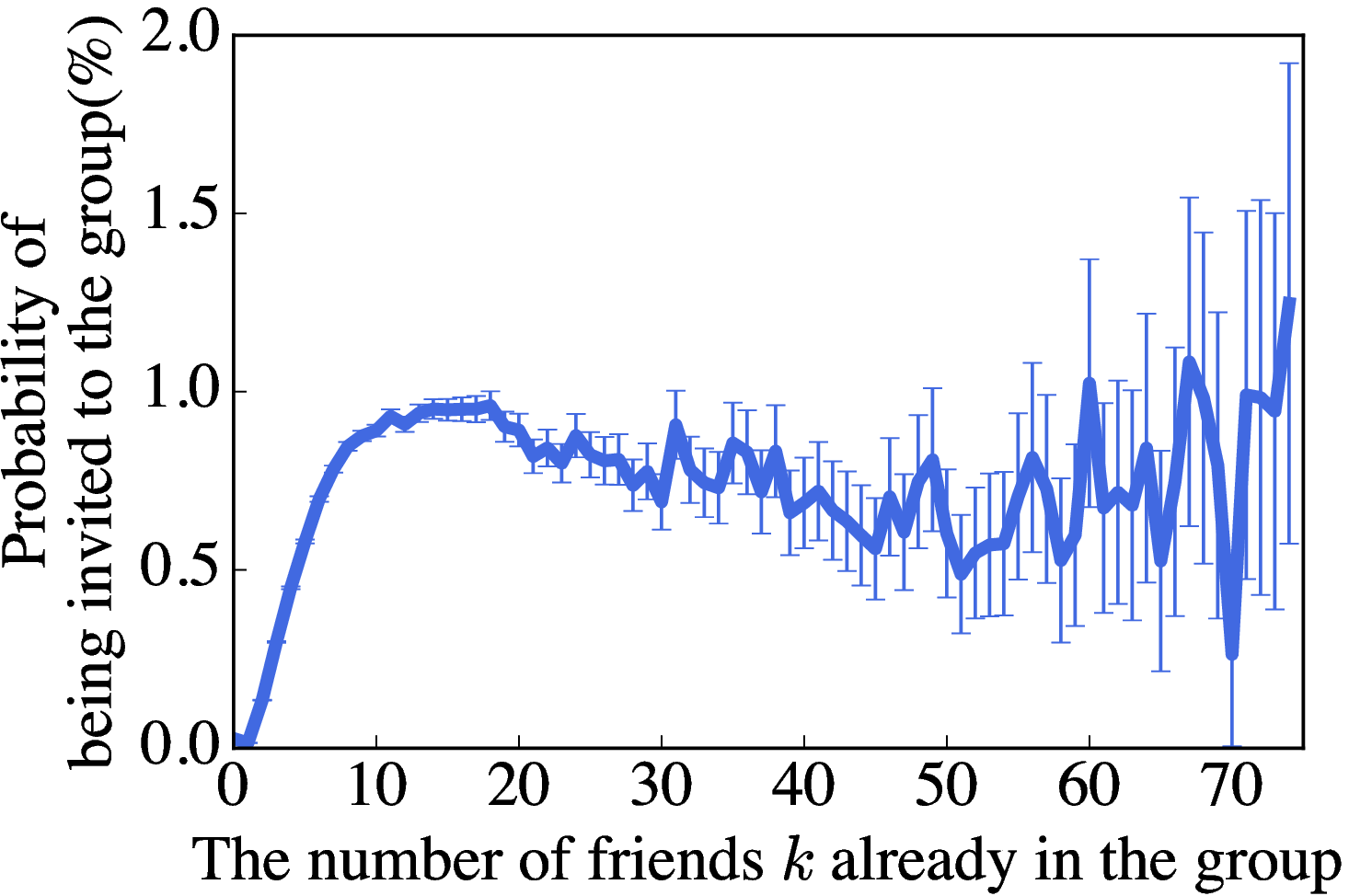}
	}
	\hspace{-0.15in}
	\subfigure[Structure diversity] {
		\label{fig:structureDiversity}
		\includegraphics[width=.78\columnwidth]{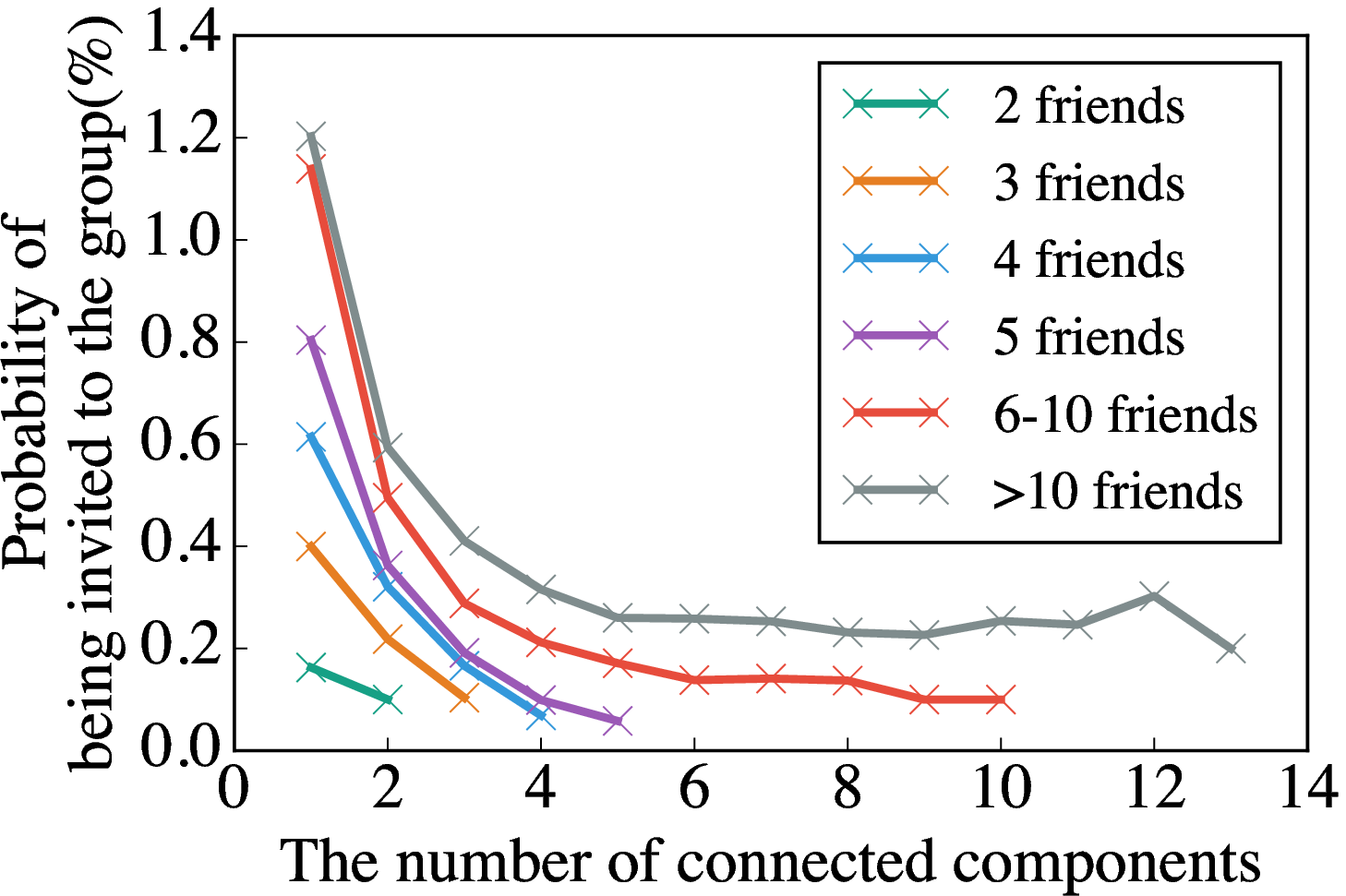}
	}
}
\vspace{-0.15in}
\caption{Local structure pattern of invitations. (a): Illustration of a potential invitee's ego networks. Give a group $\mathcal{C}$,
blue nodes represent user $v$'s friends who have been in the group, while the white nodes
denote those users in $v$'s ego networks who did not join the group.
(b): The probability $p$ of being invited to a WeChat group as a function of
the number of friends $k$ already in the group. Error bars represent 95\% confidence interval. 
(c): The effect of structure diversity. Structural diversity is represented by the number of connected components formed by the friends already in the group.
	Horizontal axis is the number of connected components formed by friends already in the group
	and the vertical axis is the probability $p$ of being invited to a WeChat group.
}
\end{figure*}

\begin{table*}[p]
\centering
\caption{List of features used in this study.}
\label{tbl:feature}
\begin{tabular}{l|l}
\hline \hline
\multicolumn{2}{c}{Group Level~(for group $\mathcal{C}$ at time $T$)} \\ \hline
\multirow{6}{*}{Group Structre}  & Number of individuals with a friend in $\mathcal{C}$~(the \textit{fringe} of $\mathcal{C}$) \\
								& Number of edges with one end in the group and the other in the fringe. \\
								& Number of edges with both ends in the group. \\
								& The number of open triads at time $T$ and at the setting up of group. \\
								& The number of closed triads at time $T$ and at the setting up of group. \\
								& Clustering coefficient. \\
\hline
\multirow{5}{*}{Cascade Tree}    & Number of Members~($|\mathcal{C}|$). \\ 
								 & Tenth quantile of members' depth in the cascade tree. \\
								 & Tenth quantile of members' subtree size. \\
								 & Number of members whose depth equal to $k$, $k=1,2,\ldots, 9$.\\
								 & The sum of the lengths of the shortest paths between all pairs of vertices, i.e., Wiener index. \\
\hline
\multirow{6}{*}{Demographics}	& Number, fraction of members from country $X$. \\
								& Number, fraction of members who stated their gender to be $X$. \\ 
								& Number, fraction of membes who stated their age to be $X$.\\
					   & Entropy of member's region~(country, province, city) distribution, e.g., $-\sum_{x\in \textrm{Countries}}P(x)\log_2{P(x)}$. \\
			  & Entropy of member's age: $-\sum_{x\in \textrm{Ages}}P(x)\log_2{P(x)}$. \\
			  & Entropy of member's gender: $-\sum_{x\in \textrm{Genders}}P(x)\log_2{P(x)}$. \\
\hline
\hline
\multicolumn{2}{c}{Inviter Level~(for member $u$ in group $\mathcal{C}$ at time $T$)} \\\hline
\multirow{4}{*}{History Behavior}		& How long has it been since $u$ joined $\mathcal{C}$. \\
										& How long has it been since $u$ invited others to $\mathcal{C}$. \\
										& The number of users that $u$ invited to $\mathcal{C}$ before time $T$. \\
										& The number of chat made by the individual. \\
\hline
\multirow{7}{*}{Local Structure}		& The number of $u$'s friends in the fringe with $\geq k$ friends in the group, $1 \leq k \leq 20, k=30, 40, 50$\\
																				  & The number and fraction of $u$'s friends in the group: $\left|\{v|v\in ego(u)\ \wedge v\in \mathcal{C}\}\right|$. \\
															 & The number and fraction of $u$'s friends not in the group: $\left|\{v|v\in ego(u) \wedge v\not\in \mathcal{C}\}\right|$. \\
													   & The number of cross group edges in $u$'s ego networks: $\left|\{(v, w)|v \in \mathcal{C} \wedge w \not\in \mathcal{C} \wedge u, w \in ego(u)\}\right|$.\\
										& The ratio of cross group edges to cross group edges that could possible exist in $u$'s ego networks. \\
										& The depth of $u$ in cascade tree. \\ \hline
\hline
\multicolumn{2}{c}{Invitee Level~(for user $u$ in the fringe of group $\mathcal{C}$ at time $T$)} \\\hline
\multirow{6}{*}{Demographics} & User $u$'s stated gender.\\
								 & User $u$'s stated age.\\
								 & User $u$'s stated region~(country, province and city). \\
								& The fration of group member who has the same gender as $u$. \\
								& The fration of group member who has the same age as $u$. \\
								& The fration of group member who has the same region~(country, province, city) as $u$. \\
\hline
\multirow{3}{*}{Local Structure}     & Number of friends already in the group. \\
																					 & Number of friend in group $\mathcal{C}$ who are classified as acitive inviters. \\
										& Number of connected components formed by friends already in the group. \\
\hline
\hline
\end{tabular}
\end{table*}

In this subsection,  we probe into the local structure of group and investigate how the membership cascade process is influenced by the structural features. Specifically, we study how the probability for a user $u$ being invited by his/her friend into a group is affected by the structure of ego networks of $u$. 

\vpara{Backstrom et al. Revisit.} In the context of online social networks, Backstrom et al. \cite{backstrom2006group} introduced the probability that an individual adopts and joins a community with the number of friends already in the group. 

As noted earlier, a significant difference in our setting of social messaging group is that new group members are ``passively invited'' to the group rather than  ``actively adopt'' the group. To this end, we calculate $P(k)$ --- the fraction of user who is invited to a community as a function of the number of $k$ of their friends who are already members, with slightly tweaking the definition in \cite{backstrom2006group}.  Specifically, we first take two snapshots of group membership, with 10 days apart. We find tuples $(u,\mathcal{C},k)$ where $u\notin \mathcal{C}$ at the time of the first snapshot and $u$ has $k$ friends in $\mathcal{C}$ at that time. We then compute $P(k)$ by looking at the fraction of tuples $(u,\mathcal{C},k)$ that $u\in \mathcal{C}$ at the time of the second snapshot.

The results for WeChat group~(see \figref{fig:kfriend}) exhibit an interesting contrast to the curves of LiveJournal and DBLP groups shown in Backstrom, Huttenlocher, Kleinberg and Lan \cite{backstrom2006group}. 
Instead of observing an increasing trend of the curve with respect to larger values of $k$,
we find a qualitatively different shape where the adoption probability suffers from a slight decrease at moderate values of $k$,
and drastic fluctuations when $k$ exceeds 40. We infer such difference is caused by the mechanism inherent to WeChat, that when a group has more than 40 users, inviting friends to join the group requires their confirmation. We leave the detailed inspection on this for future investigation. 

Furthermore, we also observe a strong evidence for the influence of structural locality. In particular when $k$ is small, the fraction of invitee~(invited to the group) with 10 friends already in the group ($k=10$) is twice as much as the fraction of invitee with 5 friends in the group ($k=5$).

\vpara{Structural Diversity.} The above analysis informs us that the number of friends in the group can affect whether a user gets invited. Yet it is unclear how it is affected by the local network structure. In this subsection, we study how  structural diversity of user $u$'s ego networks \cite{ugander2012structural} affects the probability for $u$ to be invited by friends into a group.

For a given user $u$ with $k$ friends already in a group, we measure structural diversity by counting the connected components in the networks formed by these $k$ friends. For example, as illustrated in \figref{fig:ego}, user $v$ has four friends ($A, B, C$ and $D$) already in a group. $A, B, C$ and $D$ form 3 connected components, i.e., $\{A, B\}, \{C\}$ and $\{D\}$. \figref{fig:structureDiversity} plots the curves of the probability of being invited to a group with respect to the number of connected components formed by friends already in group.
We choose the parameter $k$ (i.e., the number of friends already in group) to be 2, 3, 4, 5, 6-10 and $>10$, respectively. Interestingly, we find that given a fixed $k$, the more closely these $k$ friends are connected, the more likely $u$ will be invited to the group.

\subsection{Membership Cascade Prediction}

Now we have seen how the membership cascade process can be affected by both users' behavioral pattern as well as the local structure surrounding a user. In this subsection, we propose a prediction model by integrating a comprehensive set of behavioral and structural features. This can be used in practice to make effective inference on the membership cascade process. Specifically, given the historical behavior of group members as well as the local social structure, can we predict which members in the group are more likely to be active and invite new users to the group chat and to whom will he/she send invitations to? 

To address the issue, we separately model the inviter prediction and invitee prediction problems. 

\vspace{-.5em}
\vpara{Inviter Prediction.}
At some timestamp $T$, for a given group $\mathcal{C}$ and a specified user $u\in \mathcal{C}$, our learning task here is to predict whether $u$ will become active and invite friends to $\mathcal{C}$ during the time interval $(T,T+\Delta t]$. 

\vspace{-.5em}
\vpara{Invitee Prediction.} At some time $T$, for a given group $\mathcal{C}$ and a user $u\in fringe(\mathcal{C})$, our learning task here is to predict whether $u$ (one-hop neighbor of current members) will be invited to   $\mathcal{C}$ during the time interval $(T,T+\Delta t]$. 

For both prediction task, we construct each training example by randomly selecting $T$ from 10 minutes to 1 month, and fixing $\Delta t=1$ day. We also notice that, on average, only 5.6\% of group members have invitation action
and fewer than 1\% among users in the fringe of groups are invited, causing the number of positive examples and negative examples quite unbalanced. 
We thus down-sample \cite{kubat1997addressing} the size of negative examples and maintain a positive/negative ratio of 1:2 in our training set. 

\begin{table}[htp]
\vspace{-0.15in}
\centering
\caption{Performance of inviter/invitee prediction and 
inviter/invitee-level feature contribution analysis~(\%).}
\label{tbl:Pinviter}
\begin{tabular}{c|c|c|c|c|c}
\hline
\hline
\textbf{Task}  &\textbf{Features used} & \textbf{AUC} & \textbf{Prec.} & \textbf{Rec.} & \textbf{F1} \\
\hline
\multirow{3}{*}{Inviter} & All & \textbf{95.31} &  \textbf{85.95} & \textbf{88.39} & \textbf{87.15} \\
						 & -History Behavior & 91.52 & 82.07 & 84.31 & 83.17\\
						&-Local Structure & 93.22 & 84.50 & 87.04 & 85.75 \\
\hline
\multirow{3}{*}{Invitee} & All & \textbf{98.66} &  \textbf{54.55} & \textbf{93.47} & \textbf{68.89} \\
						& -Demographics & 98.05 & 45.76 & 94.68 & 61.70 \\
						& -Local Structure & 89.29 & 11.85 & 76.53 & 20.52\\
\hline
\hline
\end{tabular}
\vspace{-0.10in}
\end{table}

\hide{
\begin{table}[htp]
\centering
\caption{Performance of invitee prediction and 
invitee level feature contribution analysis~(\%).}
\label{tbl:Pinvitee}
\begin{tabular}{c|c|c|c|c}
\hline
\hline
\textbf{Features used} & \textbf{AUC} & \textbf{Prec.} & \textbf{Rec.} & \textbf{F1} \\
\hline
All & \textbf{98.66} &  \textbf{54.55} & \textbf{93.47} & \textbf{68.89} \\
-Demographics & 98.05 & 45.76 & 94.68 & 61.70 \\
-Local Structure & 89.29 & 11.85 & 76.53 & 20.52\\
\hline
\hline
\end{tabular}
\end{table}
}

We incorporate both group-level and inviter-level features seen in Table \ref{tbl:feature} for training the inviter model; and use instead the group-level and invitee-level features for invitee model.
We further train the dataset using support vector machine (SVM)~\cite{fan2008liblinear} with 10-fold cross validation.
The prediction performance results are shown in Table \ref{tbl:Pinviter}. We see that our model is quite effective, with AUC of 95.31\% in predicting inviter, and an AUC of 98.66\% in predicting invitee. We further investigated how each set of features affects the training performance by  considering only one at a time. Quite interestingly, we find that historical behavioral features can be important factors in the task of predicting inviter; while local structural features are the dominant ones in predicting invitee. The information of demographics exert little affect on the performance of predicting invitee, which implies that our model can be generalizable without requiring user-specific attributes.

\section{conclusion}
\label{sec:conclusion}

\vpara{Summary.} In this paper, we studied  the formation and evolution of chat groups in the context of social messaging --- their lifecycles, the change in their underlying structures over time, and the diffusion processes by which they develop new members. We use a large collection of anonymized data from WeChat group messaging platform, providing analysis on dynamics of millions of groups by keeping track of their emergence, growth and demise over time. We discovered a strong dichotomy of groups existed in terms of their lifecycle, and defined two types of groups accordingly: long-term and short-term groups. First, we developed an effective separability model by taking into account a broad range of group-level features, showing that long-term and short-term groups are inherently distinct. We also found that the lifecycle of messaging groups is largely dependent on their social roles and functionalities in users' daily social experiences and specific purposes. Specifically, event-driven groups in general have a shorter lifespan as apposed to those friendship groups serving for frequent catching-up purpose. Given the strong separability between the long-term and short-term groups, we further addressed the problem of early prediction in group longevity, and demonstrated that strong prediction results can be obtained even with a group's history up to one day.

In addition to modeling the growth and evolution from a group-level perspective, we also investigated the individual-level attributes of group members and study the diffusion process by which groups gain new members. We developed a membership cascade process model in which we consider users’ historical engagement behavior as well as the local social network structure that users embedded in. We demonstrated the effectiveness  by achieving AUC of 95.31\% in the inviter prediction model using all features, and an AUC of 98.66\% in the invitee prediction model.


\vpara{Future Research.} Ours findings raise many important open questions that would be interesting to take into account in future research. First, our design of membership cascade model can be used for group member recommendation, and may be potentially integrated into current WeChat platform. This can motivate research on conducting online experiments and investigating whether users are likely to adopt the group member recommendations, and under what circumstances. Such studies will also lead to design of better group chat platforms and engage users more effectively.


\vpara{Acknowledgements.}  We thank Chenhao Tan and Tracy Xiao Liu for their comments.
Jiezhong Qiu and Jie Tang are supported by 
863~(No. 2014AA015103),
973~(No. 2014CB340506,
No. 2012CB316006),
NSFC~(No. 61222212), and NSSFC~(No. 13\&ZD190).
Yixuan Li and John E. Hopcroft are supported by the US Army Research Office W911NF-14-1-0477.
Qiang Yang is supported by 
973~(No. 2014CB340304).
This work is also supported by research fund of Tsinghua-Tencent Joint Laboratory.


\newpage

\bibliographystyle{abbrv}
\bibliography{sigproc}
\normalsize
\end{document}